\documentclass[11pt] {article}
\usepackage[a4paper,margin=1.1in]{geometry}
\usepackage{lineno}

\usepackage[T1]{fontenc} %
\usepackage[normalem]{ulem} %

\usepackage[usenames,dvipsnames,svgnames,x11names]{xcolor}

\usepackage{cite}

\usepackage{balance}

\usepackage{listings}

\lstdefinelanguage{XML}
{
basicstyle=\ttfamily\footnotesize,
  morestring=[b]",
  moredelim=[s][\bfseries\color{Maroon}]{<}{\ },
  moredelim=[s][\bfseries\color{Maroon}]{</}{>},
  moredelim=[l][\bfseries\color{Maroon}]{/>},
  moredelim=[l][\bfseries\color{Maroon}]{>},
  morecomment=[s]{<?}{?>},
  morecomment=[s]{<!--}{-->},
  commentstyle=\color{gray},
  stringstyle=\color{blue},
  identifierstyle=\color{red}
}

\usepackage{moreverb}

\usepackage[nounderscore]{syntax}

\usepackage[pdftex]{graphicx}
\graphicspath{{./figures/}}
\DeclareGraphicsExtensions{.pdf}

\usepackage[cmex10]{amsmath}
\usepackage{amssymb}
\usepackage{mathtools}
\usepackage{amsthm}
\usepackage{amsfonts}
\usepackage{gensymb}

\usepackage{subfig} %
\usepackage{algorithmicx}
\usepackage{algpseudocode}
\usepackage[ruled]{algorithm}
\definecolor{light-gray}{gray}{0.75}
\algrenewcommand{\algorithmiccomment}[1]{\hskip3em{{\footnotesize \textcolor{light-gray}{$\blacktriangleright$}}} #1}

\usepackage{multirow} %
\usepackage{rotating} %
\usepackage{booktabs} %
\usepackage{colortbl} %
\usepackage{tablefootnote} %

\usepackage{array}
\newcolumntype{L}[1]{>{\raggedright\let\newline\\\arraybackslash\hspace{0pt}}m{#1}}
\newcolumntype{C}[1]{>{\centering\let\newline\\\arraybackslash\hspace{0pt}}m{#1}}
\newcolumntype{R}[1]{>{\raggedleft\let\newline\\\arraybackslash\hspace{0pt}}m{#1}}

\usepackage[pdftex,colorlinks=true,urlcolor=blue,citecolor=blue]{hyperref}

\usepackage{xspace}

\usepackage{enumitem}

\usepackage[usenames,dvipsnames,svgnames,x11names]{xcolor}

\newcommand{\delc}[1]{ {\textcolor{gray} {\sout{#1}} }}

\renewcommand{\delc}[1]{} 

\usepackage{blindtext}

\newcommand{\at}{ATLAS\xspace}
\newcommand{\ta}{Taurus\xspace}

\newcommand{\M}{\textsc{M}}
\newcommand{\B}{\textsc{B}}

\newcommand{\NS}{\texttt{NOT\_STARTED}\xspace}
\newcommand{\OG}{\texttt{ON\_GPU}\xspace}
\newcommand{\IB}{\texttt{IN\_BUFFER}\xspace}
\newcommand{\EV}{\texttt{EVICTED}\xspace}
\newcommand{\CP}{\texttt{COMPLETED}\xspace}

\def\orcid#1{\kern .08em\href{https://orcid.org/#1}{\includegraphics[keepaspectratio,width=0.7em]{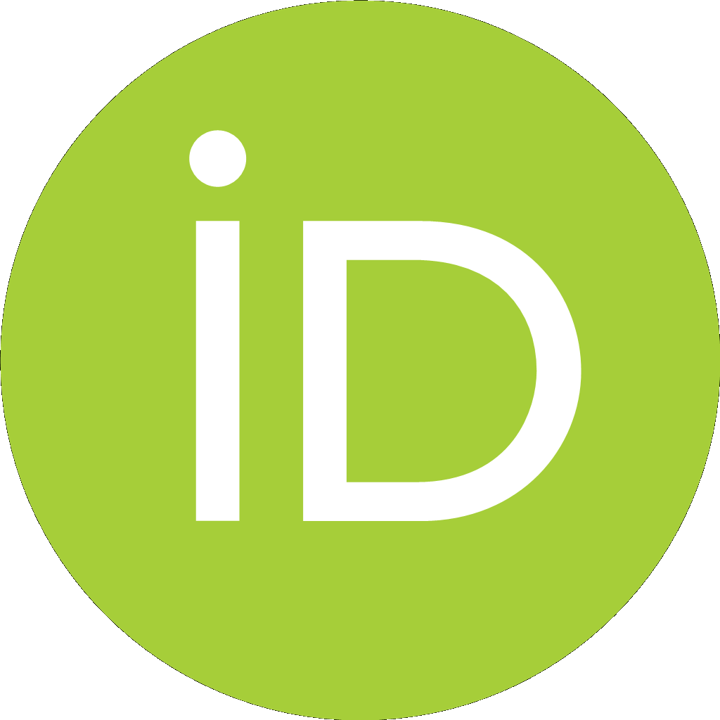}}}
    
\begin{document}

\title{\ta\thanks{\ta is the constellation that houses the star, \textit{Atlas}, reflecting its evolution from our prior out-of-core GNN inference framework, \at~\cite{atlas}.}~: Accelerating Out-of-Core Graph Neural Network Inference on Billion-Scale Graphs~\thanks{~Extended full-length version of paper that appeared at HPDC 2026: \textit{``\at: Efficient Out-of-Core Inference for Billion-Scale Graph Neural Networks'', Pranjal Naman and Yogesh Simmhan, in the 35th ACM International Symposium on High-Performance Parallel and Distributed Computing (HPDC), 2026}. DOI: \url{https://doi.org/10.1145/3806645.3807597}}
}

\author{Pranjal Naman\orcid{0009-0000-9912-9522} and Yogesh Simmhan$^1$\orcid{0000-0003-4140-7774}\\~\\
\em Department of Computational and Data Sciences (CDS),\\
\em Indian Institute of Science (IISc),\\
\em Bangalore 560012 India\\~\\
\texttt{Email:\{pranjalnaman, simmhan\}@iisc.ac.in}
}
\date{}
\maketitle

\begin{abstract}
Graph Neural Network (GNN) inference on billion-scale graphs is challenging due to the large memory footprint of features and embeddings and high disk I/O costs in out-of-core settings. Existing distributed GNN systems incur high communication times and infrastructure costs while disk-based GNN systems are primarily tailored to training and experience massive wasted reads during inference on the entire graph. We present \textsf{\ta}, a \textit{single-machine} system for GNN inference on graphs that do not fit in RAM, supporting both \textit{exact} full-graph inference and fanout-sampled inference.
To avoid random and repeated feature gathers, \ta reformulates layer-wise inference as source-centric broadcasts over sequential SSD scans, backed by a pipelined GPU--CPU--SSD hierarchy, topology-aware reordering, pending-message eviction and a GPU-resident store for high-degree vertices.
It further uses non-buffered sequential reads and GPU-backed writes to reduce page-cache pollution, host-memory pressure and write overheads.
On out-of-core graphs with up to $269\M$ vertices, $4\B$ edges, and $514$~GiB of features, \ta outperforms the strongest layer-wise baseline, DGI, by $7$--$25\times$, and vertex-wise baselines by $40$--$140\times$.
\end{abstract}

\section{Introduction}\label{sec:intro}

Graph Neural Networks~(GNNs) are an effective tool for learning representations from graph data, capturing both topology and associated features~\cite{kipf2017gcn, hamilton2017sage}. This makes them adept at performing a variety of tasks like detecting fraud in transaction networks~\cite{dou2020enhancing,bharadwaj2026npci}, predicting traffic flows and signaling in Intelligent Transportation Systems~(ITS)~\cite{derrow2021eta,simmhan2026scale}, and making e-commerce recommendations~\cite{aligraph}.
Most GNNs follow a \textit{message-passing} paradigm, where each vertex iteratively \textit{gathers} and \textit{aggregates} information from its neighbors while applying a neural-network transformation at every layer.
This iterative neighborhood aggregation incurs significant computational and memory overhead due to irregular graph accesses and repeated neural network computations.

\subsubsection*{Motivation}
Given these memory and computational costs, optimizing inference is critical for real-world deployments. GNNs are often deployed on evolving graphs requiring predictions to be periodically refreshed~\cite{inkstream, ripple}. 
While this refresh is typically not latency-sensitive, it must complete within practical timeframes, i.e., hours rather than days, using either exact or sampled inference depending on application needs.
This is further complicated by the scale of real-world graphs, often containing millions--billions of vertices and edges, e.g., fintech transaction networks for fraud detection and social or e-commerce graphs for recommendation tasks. Although graph topology \textit{may} fit in the RAM of a single machine, vertex features/embeddings often dominate the memory footprint. E.g., the \textit{IGB-Full} citation graph~\cite{igb}~($269\M$ vertices, $4\B$ edges, $1024$ FP16 features) requires \textit{514~GiB of RAM}~(Table~\ref{tab:datasets}). To circumvent this, 
distributed GNN systems partition graphs across multiple servers and perform training or inference collaboratively on distributed subgraphs, incurring high infrastructure and network costs.
Recent inference systems further optimize performance through probabilistic caching of remote features and communication-aware graph and feature partitioning~\cite{kaler2023communication,chen2025deal}.

\textit{Disk-based GNN training} has emerged as a promising alternative~\cite{diskgnn, caliex, capsule, outre, ginex}, along the lines of prior \textit{Out-of-Core~(OOC)} disk-based parallel graph processing~\cite{vora2019lumos,roy2013x}. These focus on efficient data layouts and intelligent caching strategies to fully utilize memory, disk capacity, and bandwidth on a single machine. This is all the more relevant given the lower latency and higher bandwidth of Solid State Disks~(SSDs) with capacities of 2~TiB+ common even for prosumer disks, at a much lower price point than RAM.

While disk-based OOC GNN training has been extensively studied, inference presents distinct challenges that remain largely unexplored. \textit{(i) Working-Set Amplification During Inference:} Existing OOC training systems optimize \textit{data transfer}~\cite{hyperion, gids}, \textit{data organization} through reordered movement and multi-tier caching~\cite{diskgnn, ginex, capsule, outre}, and \textit{storage layouts} tailored for efficient training~\cite{diskgnn}.
While inference, at first glance, appears to be just the forward-pass phase of training, these are fundamentally different workloads. Training operates on a small labeled subset of vertices (e.g., $\approx1\%$ in \textit{OGBN-Papers100M}~\cite{ogb}), whereas inference computes embeddings for the entire graph. As a result, OOC training optimizations that exploit a limited working set, such as computation-graph precomputation~\cite{diskgnn, capsule}, do not readily apply because materializing computation graphs for all vertices is prohibitively expensive.
\textit{(ii) Sampling Effects on Inference: } Training typically employs \textit{neighborhood sampling}~\cite{hamilton2017sage} to bound memory usage by aggregating over only a subset of neighbors. 
Inference may require \textit{exact full-neighborhood aggregation} for accuracy-critical domains such as fintech and ITS, where sampling can introduce non-deterministic predictions~\cite{inferturbo, kaler2022accelerating, ripple}; other applications may tolerate \textit{fanout-sampled} inference.
In both modes, disk-resident inference still streams vertex features or intermediate embeddings across layers, and sampling mainly reduces message propagation.
E.g., training a 2-layer GNN on \textit{IGB-Large} with $1\%$ labeled vertices touches only $\approx8\%$ of all vertices per epoch~\cite{naman2025gpu}.
Consequently, OOC GNN inference cannot directly inherit training optimizations.
It must control read amplification, active-state residency and output materialization. 

\begin{figure}
    \centering
    \subfloat[Random and repeated accesses\label{subfig:motivation-1a}]{
        \includegraphics[width=0.25\columnwidth]{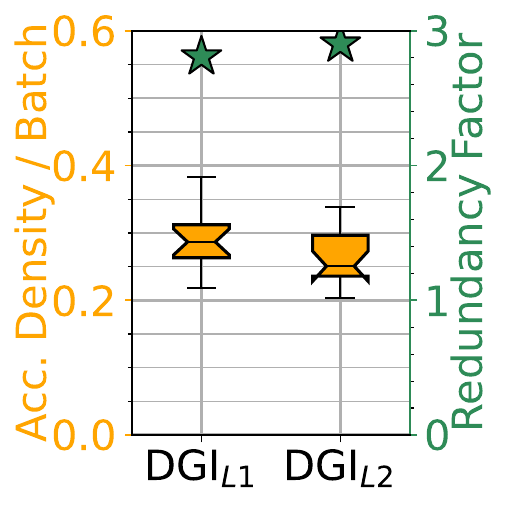}
    }\qquad\qquad
    \subfloat[Read amplification\label{subfig:motivation-1b}]{
        \includegraphics[width=0.36\columnwidth]{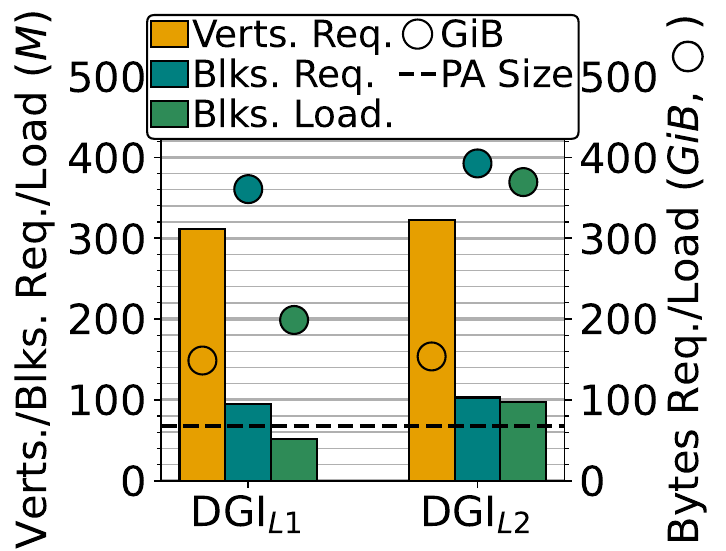}
    }
    
    \caption{Challenges limiting out-of-core GNN inference on \textit{Papers}~\cite{ogb}. (a) Random and repeated accesses quantified by access density~(left Y axis) and redundancy factor (\textit{stars}, right Y axis). (b) Read amplification: requested vertices/blocks and loaded blocks (\textit{bars}, left Y axis), requested/loaded bytes (\textit{markers}, right Y axis), and \textit{Papers} feature size (\textit{dashed line}).}
    \label{fig:motivation-1}
\end{figure}

\subsubsection*{Challenges}

Existing out-of-core~(OOC) GNN inference typically uses either \textit{vertex-wise}~\cite{wang2019deep, ginex, diskgnn} or \textit{layer-wise}~\cite{dgi} gather-based execution.
Vertex-wise methods~(e.g., DGL~\cite{wang2019deep}) recursively aggregate $k$-hop neighborhoods for each batch, while layer-wise methods~(e.g., DGI~\cite{dgi}) compute embeddings for all vertices one layer at a time to eliminate redundant neural-network computations. Despite these differences, both suffer severe I/O bottlenecks on large disk-resident graphs. 
We demonstrate these inefficiencies using DGI~(our strongest layer-wise baseline) for $2$-layer GraphConv~(\textit{fanout=10})~\cite{kipf2017gcn} sampled inference on the RCMK-reordered \underline{PA}pers graph (Table~\ref{tab:datasets}) on a $32$~GiB RAM machine:

\begin{enumerate}[leftmargin=0pt,itemindent=20pt,label={\em(\arabic*)},topsep=0pt,listparindent=\parindent]
\item \textit{Random Access}: Neighbors required for inference are scattered across the feature store (Fig.~\ref{fig:example}b, \textit{top}). \textit{Access density} -- the ratio of neighbors fetched per batch to the ID range they occupy -- for DGI stays $\approx 0.2$--$0.3$ across both layers (Fig.~\ref{subfig:motivation-1a}, \textit{box plots}, left Y axis), well below $1$ even with RCMK reordering ($1=$ contiguous accesses).
\item \textit{Repeated Access}: Neighborhoods fetched across batches have a high overlap leading to repeated fetches (Fig.~\ref{fig:example}b, \textit{center}). \textit{Redundancy factor} -- the total vertex fetches divided by unique vertices processed overall -- is $\approx 3\times$ for both layers using DGI (Fig.~\ref{subfig:motivation-1a}, \textit{stars}, right Y axis).
\item \textit{Read Amplification}: Random and repeated accesses translate
directly into excess disk I/O; since storage is accessed at block
granularity (e.g., 4~KiB), scattered reads fetch data that remains unused
(Fig.~\ref{fig:example}b, \textit{bottom}). Across both layers, DGI
requests $\approx750$~GiB of feature blocks (Fig.~\ref{subfig:motivation-1b}, \textit{teal circles}) against only $\approx300$~GiB of features actually needed by the requested vertices (\textit{orange circles}). Although the OS page cache absorbs a portion of this, physical reads still total $568$~GiB (\textit{green circles}) -- nearly $1.9\times$ the useful bytes requested, and over $8.35\times$ of PA's $68$~GiB total size.
\end{enumerate}

\begin{figure}[t]
    \centering
    \subfloat[PA/GCN2\label{subfig:motivation-2a}]{
        \includegraphics[width=0.3\columnwidth]{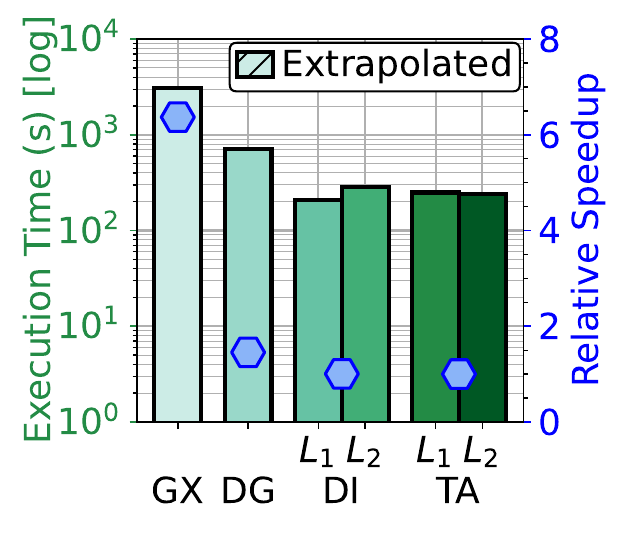}
    }\qquad\qquad
    \subfloat[MA/GCN2\label{subfig:motivation-2b}]{
        \includegraphics[width=0.3\columnwidth]{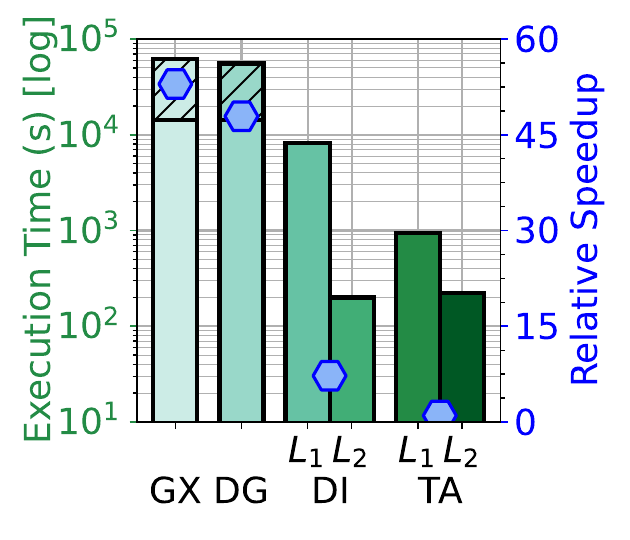}
    }
    \caption{Inference time (left Y axis, \textit{bars}) and speedup relative to \ta (right Y axis, \textit{markers}) for 2-layer \textit{GraphConv} inference with \textit{fanout=10} on disk-resident graphs~(topology+features) reported for \underline{G}ine\underline{X}, \underline{DG}L, \underline{D}G\underline{I}, and \textsf{\underline{TA}urus} (ours) on \textit{\underline{PA}pers} and \textit{\underline{MA}G-Cites} using a 5090 GPU.}
    \label{fig:motivation-2}
\end{figure}

\textit{The common root cause is destination-centric gathering.} Since each destination independently pulls neighbor embeddings, systems repeatedly move the same source data, lose sequentiality and amplify disk traffic even with graph reordering.
We also \textit{empirically demonstrate} these challenges in Fig.~\ref{fig:motivation-2} by comparing our proposed \ta with three baselines: \textit{DGI}~\cite{dgi}~(layer-wise inference), and \textit{Ginex}~\cite{ginex} and \textit{DGL}~\cite{wang2019deep}~(training frameworks adapted for vertex-wise inference), using sampled 2-layer GraphConv inference~\cite{kipf2017gcn} over \underline{PA}pers and \underline{MA}G-Cites (Table~\ref{tab:datasets}), on a \textit{GPU workstation} with $128$~GiB RAM, RTX $5090$ GPU and $2$~TiB SSD~(\S~\ref{subsec:eval-setup}).
Both DGL~\cite{wang2019deep} and Ginex~\cite{ginex} are unable to complete the inference even within a $4$~h time budget on MA, taking an extrapolated $\approx16$~h~(Fig.~\ref{subfig:motivation-2b}, left Y axis), while DGI~\cite{dgi} took $\approx2.5$~h. In contrast, our \ta framework completes this in $<0.5$~h for both layers.

\subsubsection*{Proposal}
To address these challenges, we leverage a key insight: 
\textit{layer-wise GNN inference can be reformulated as source-centric broadcasts, enabling sequential disk access instead of repeated random gathers.}
This reduces read amplification, but na\"{i}ve broadcasts merely shift the bottleneck to partial-state memory pressure and random output writes.
Thus, \ta must preserve sequential reads while bounding active states and avoiding random-write amplification.

\subsubsection*{Contributions}

We present \textbf{\ta}, a disk-based GNN inference framework for billion-scale graphs on a single workstation. \ta supports exact full-graph and fanout-sampled inference by replacing destination-centric gathers with source-centric broadcasts, combined with tiered GPU--RAM--SSD aggregation, topology-aware reordering and eviction, and pipelined I/O, aggregation, GPU compute and output.
Specifically, we make the following contributions:

\begin{enumerate}[leftmargin=*]
\item \textit{Broadcast-based inference model.}
\ta introduces broadcast-based layer-wise inference that reads features and embeddings sequentially, reducing read amplification relative to gather-based approaches. It supports both exact full-graph and fanout-sampled inference while preserving their respective GNN semantics.

\item \textit{GPU-enhanced tiered runtime.}
\ta employs a pipelined GPU--RAM--SSD hierarchy that overlaps I/O, aggregation, transformation, and output, reducing memory pressure and data movement overheads while maintaining high throughput under constrained memory.

\item \textit{Architecture support and topology-aware execution.}
We extend broadcast inference to Graph Attention Networks and memory-efficient GraphSAGE, and introduce a topology-aware graph reordering strategy that reduces partial-state residency and evict-reload cycles during execution.

\item \textit{Comprehensive evaluation.}
We evaluate \ta on billion-scale citation and social-network graphs with feature sizes above $500$~GiB under exact and sampled inference. We show substantial reductions in disk traffic and runtime over DGL GraphBolt-backed and out-of-core baselines, with ablations of key runtime components.
\end{enumerate}

This article extends our previous conference work \at~\cite{atlas}, which introduced broadcast-based layer-wise GNN inference using a RAM--SSD hierarchy and pipelined execution. \ta extends it with: (1) GPU-resident aggregation and GPU-backed output materialization to reduce host-memory pressure and write overheads; (2) A topology-aware reordering objective that targets partial-state residency and eviction--reload cycles; (3) Support for fanout-sampled inference, memory-efficient GraphSAGE and Graph Attention Networks; and (4) Additional billion-scale datasets, a DGL GraphBolt baseline and expanded ablations.

\section{Background}\label{sec:background}
\begin{figure}[t]
    \centering
    \includegraphics[width=0.6\linewidth]{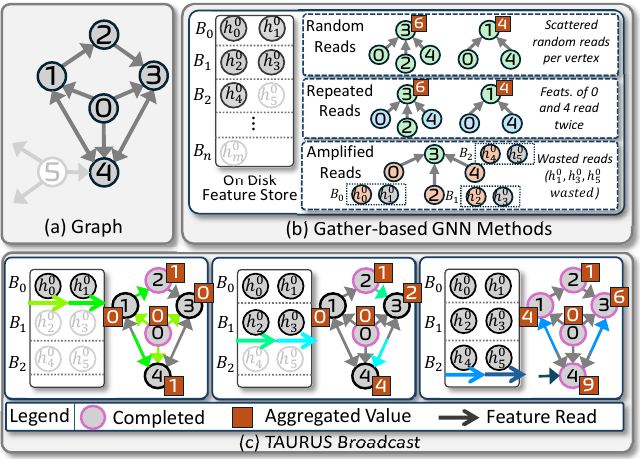}
    \caption{Gather-based versus broadcast-based execution for one GNN layer.}
    \label{fig:example}
\end{figure}

\subsection{GNN Training and Inference}
A GNN layer gathers and aggregates the embeddings of a vertex $u$'s in-neighbors~($N^-(u)$) to produce an intermediate representation $x_u^l$ (Eqn.~\ref{eq:gnn-1}), which is transformed by a learnable \textsc{Update} function and non-linear activation $\sigma(\cdot)$ to generate $h_u^l$ (Eqn.~\ref{eq:gnn-2}). Training repeats this for $L$ layers in the forward pass and then updates parameters through backpropagation.
\begin{align}
    x^l_u =&~ \textsc{Aggregate}^l(\{h^{l-1}_v, v \in N^-(u)\}) \label{eq:gnn-1}\\
    h^l_u =&~ \sigma(\textsc{Update}^l(h^{l-1}_u, x^l_u)) \label{eq:gnn-2}
\end{align}
Aggregating all in-neighbors recursively leads to \textit{neighborhood explosion} and out-of-memory (OOM) errors~\cite{hamilton2017sage}. To mitigate this, training typically employs neighborhood sampling, where only a subset of neighbors is aggregated at each hop~\cite{hamilton2017sage}.

In contrast, GNN inference requires only the forward pass and may use either exact full-neighborhood aggregation for deterministic embeddings~\cite{inferturbo, kaler2022accelerating, ripple} or fanout sampling when approximation is acceptable. Full-neighborhood \textit{vertex-wise} inference suffers from neighborhood explosion and redundant computation, motivating \textit{layer-wise} inference~\cite{dgi}, which materializes embeddings for all vertices one layer at a time. \ta adopts and optimizes layer-wise inference for out-of-core execution.

\subsection{Gather-based Execution Model}
Most GNN systems implement message passing as \textit{gather-based} execution: a destination vertex $u$ retrieves in-neighbor embeddings $\{h_v^{l-1}\mid v\in N^-(u)\}$ to compute $h_u^l$. When embeddings reside on disk, these destination-centric gathers cause irregular and repeated accesses (Fig.~\ref{fig:example}).
Message-passing execution consists of \textit{sample}, \textit{gather}, \textit{transfer} and \textit{compute}. \textit{Sampling} selects a subset of in-neighbors to limit neighborhood explosion~\cite{hamilton2017sage}; exact inference instead uses all in-neighbors. \textit{Gather} retrieves selected or full-neighborhood embeddings, \textit{transfer} moves them to the GPU, and \textit{compute} applies aggregation followed by neural-network transformations.
Popular architectures such as GraphConv~\cite{kipf2017gcn}, GraphSAGE~\cite{hamilton2017sage}, GIN~\cite{gin} and GAT~\cite{gat} follow this paradigm. \ta targets their out-of-core inference.

\subsection{Layer-wise Inference}
A common inference approach reuses training code with the backward pass disabled~(\textit{vertex-wise inference})~\cite{wang2019deep}. Under full-neighborhood aggregation, overlapping neighborhoods cause memory blowup and redundant computation. Layer-wise inference instead materializes embeddings for all vertices once per layer and reuses them in later layers~\cite{dgi, ripple, inkstream}.
However, layer-wise execution does not eliminate redundant data movement. In OOC settings, gather-based execution still makes each destination independently fetch in-neighbor embeddings, so read volume grows with propagated messages ($\approx$(sampled) edges) rather than unique vertices.
Graph reordering~\cite{chan1980linear, dgi} improves locality but cannot remove these repeated gathers. Since inference is often lightweight, OOC execution becomes I/O-bound, and repeated layer-wise access patterns amplify cumulative read traffic.

\section{System Design}\label{sec:sysdesign}
\begin{figure}[t]
    \centering
    \includegraphics[width=0.7\linewidth]{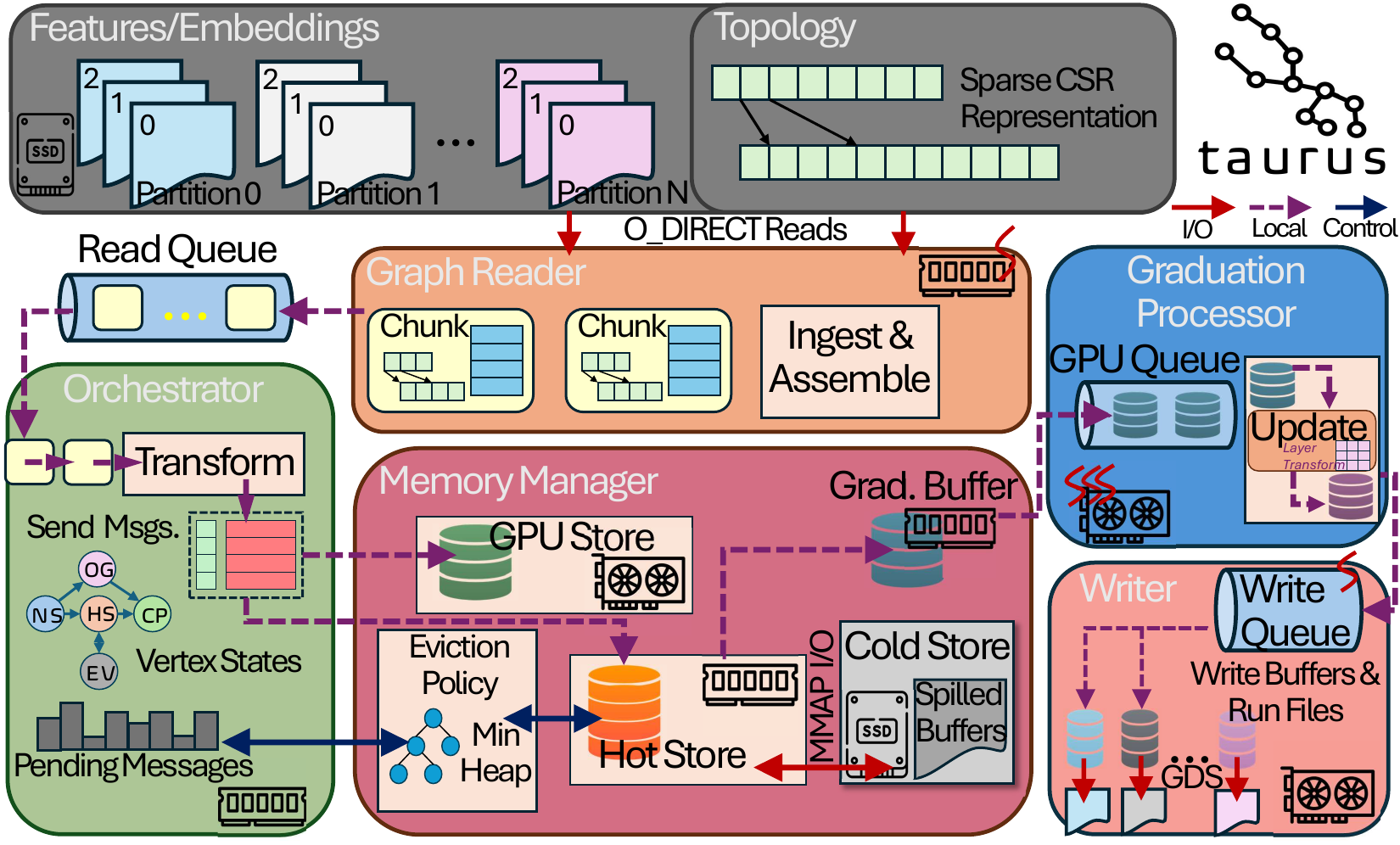}
    \caption{\ta Architecture: Sequential graph/embedding reads, tiered aggregation, GPU transformation \& run-file output.}
    \label{fig:atlas-arch}
\end{figure}

\subsection{Broadcast Execution Model and Challenges}
\ta replaces destination-centric gathers with source-centric broadcasts, streaming each source feature/embedding once per pass in vertex order (Fig.~\ref{fig:example}b), for both full-neighborhood and sampled message sets.
In contrast, gather-based inference performs scattered reads, repeatedly fetches shared source features, and loads unused records from block-granularity storage.
Broadcast is semantically equivalent to gather for the same message set, \textsc{Aggregate}, and \textsc{Update} functions, but realizing it for OOC graphs introduces several challenges.

\begin{enumerate}[leftmargin=0pt,itemindent=20pt,topsep=0pt,listparindent=\parindent,] 
\item {\em Bounded memory.} 
Broadcast execution eliminates repeated feature reads but creates many partially aggregated destination states per layer. For large graphs, these states cannot be fully materialized in memory and must be managed under a fixed budget.

\item {\em Vertex ordering.} 
Source-vertex order determines when destinations become active and complete. Poor ordering lengthens partial-state lifetimes, increasing RAM pressure and SSD spills.

\item {\em Output materialization.} 
Vertices complete only after receiving all messages, so completion order is not sequential. Writing outputs immediately can therefore replace random-read amplification with random-write amplification.

\item {\em I/O efficiency and overlap.}
Topology, embeddings, aggregation, transformation and output must be organized and pipelined so that sequential I/O, compute and writes do not stall each other.
\end{enumerate}

\subsection{Data Layout}\label{subsec:sys-data-layout}
The on-disk layout is central to OOC performance. Since \ta broadcasts along vertex \textit{out-edges}, it stores topology in Compressed Sparse Row (CSR) format~(Fig.~\ref{fig:atlas-arch}, top), requiring $\mathcal{O}(|V|+|E|)$ space and enabling sequential source-vertex scans via file offsets by the \textit{graph reader}.

For features and intermediate embeddings, vertices do not complete in vertex-ID order under broadcast execution. A dense in-memory embedding array is infeasible, while dense on-disk placement causes costly random writes. Externally sorting completed embeddings would add large temporary storage and I/O.

To address these challenges, \ta partitions the vertex-ID space into fixed ranges and maintains embeddings in sorted order within each range~(Fig.~\ref{fig:atlas-arch}, top, \textit{Partition 0--N}). Since an individual range may still exceed available memory, each partition is materialized as multiple sorted \textit{run files}, generated by sorting buffered embeddings in-memory and flushing them sequentially to disk. 
Merging spilled run files would require another large multi-way merge. Instead, \ta leaves runs unmerged and lets the \textit{graph reader} reconstruct a sequential view on demand.

\subsection{Graph Reader}\label{subsec:sys-reader}
\ta uses a partially sequential \emph{graph reader} that streams vertex embeddings once per pass~(Fig.~\ref{fig:atlas-arch}, \textit{orange}).
Input and output embeddings are stored as \emph{run files} of sorted $(\texttt{vertex\_id},\texttt{embedding})$ records within contiguous vertex-ID ranges. Each run $f$ stores IDs $I_f$, feature matrix $X_f$, and $n_f$ records, indexed by its ID range $(I_f[0], I_f[n_f-1])$. 
Multiple run files may exist within a partition. 
The graph reader exposes a \textit{chunk}-based iterator over contiguous vertex ranges. Each chunk contains CSR out-neighbors/offsets and embeddings in vertex order, and is enqueued into a \textit{reader queue} for downstream processing. For chunk size $C$ and feature size $F$, a chunk contains $\lfloor C/F \rfloor$ vertices. Chunks are partitioned by feature bytes rather than edge count. So high-degree vertices increase per-chunk edge work but not feature-read ordering.

For a chunk spanning vertex IDs $[s,e)$, the reader identifies overlapping runs. For each run $f$, two binary searches compute $\ell_f=\min\{i \mid I_f[i]\ge s\}$ and $r_f=\min\{i \mid I_f[i]\ge e\}$, yielding rows $[\ell_f,r_f)$ 
within the chunk range.
It issues one aligned \textit{pread} per overlapping run using direct I/O (\texttt{O\_DIRECT}), bypassing the OS page cache for single-pass embedding scans. Sorted rows from different runs are merged by vertex ID to reconstruct the chunk matrix. This \textit{merge-on-read} avoids an external merge sort over all output embeddings at each layer. Run file descriptors are opened lazily, and a dedicated reader thread overlaps I/O with computation.

\subsection{Orchestrator}\label{subsec:sys-orch}
The \textit{orchestrator} preserves GNN semantics under broadcast execution~(Fig.~\ref{fig:atlas-arch}, \textit{green}). It consumes reader chunks from the \textit{reader queue}~(\S~\ref{subsec:sys-reader}), initializes the \textit{memory manager} and \textit{graduation processor}, and maintains compact $\mathcal{O}(|V|)$ arrays for each vertex's pending-message count and execution state.

For a GNN layer with mean aggregation, instead of gathering neighbors at destination $v$, the orchestrator emits one normalized message per propagated edge $(u,v)$: 
$m_{u\rightarrow v}^{(l)}=\frac{1}{|\mathcal{M}(v)|}h_u^{(l)}$, where $\mathcal{M}(v)$ is the full or sampled in-neighbor set propagated for $v$.
It sends $\langle v,m_{u\rightarrow v}^{(l)},s_v\rangle$ to the memory manager, where $s_v$ is $v$'s current state.

The orchestrator tracks partial aggregation using the state machine in Fig.~\ref{fig:atlas-arch}. Vertices start in \NS, move to \IB when their first message creates a hot-store state, transition to \EV if spilled to the SSD cold store, or to \OG if pinned in the GPU store (e.g., high in-degree vertices). After all expected messages arrive, they enter \CP and become eligible for graduation. Valid transitions are \NS $\rightarrow$ \IB/\OG, \IB $\rightarrow$ \EV, \EV $\rightarrow$ \IB, and \IB/ \OG $\rightarrow$ \CP; \OG vertices are pinned.

\subsection{Memory Manager}\label{subsec:sys-mem-manager}
The memory manager maintains partial aggregation states under a fixed (configurable) RAM budget~(Fig.~\ref{fig:atlas-arch}, \textit{red}) using a three-tier GPU--RAM--SSD hierarchy: GPU store, host-memory hot store and SSD-backed cold store. High-traffic states are pinned in GPU memory, active states occupy the hot store, and overflow states spill according to the eviction policy.

\subsubsection{GPU Store}
The highest tier is a \textit{GPU store} for high in-degree vertices. Since power-law graphs route many messages to a small hub set, the memory manager pins their aggregation buffers~(\OG) in VRAM during initialization. Messages to these vertices bypass the hot store and accumulate directly in VRAM, reducing RAM pressure.

\subsubsection{Hot Store} 
The \textit{hot store} is a fixed-size host-memory slot array, with each slot holding one active vertex's partial aggregation state~(\IB). Messages accumulate through a vertex-to-slot map; slots are allocated on entry to \IB and released at \CP. When full, selected states are evicted to the SSD-backed \textit{cold store}. Since aggregation proceeds only for states in the hot or GPU store, evicted states must be reloaded before receiving further messages, and the memory manager reports all state transitions to the orchestrator.

\subsubsection{\ta Eviction Policy}
When the hot store is full, the eviction policy selects spill victims. Random or recency-based choices may repeatedly evict vertices far from completion, causing eviction--reload cycles and extra SSD I/O. \ta instead evicts vertices with the \textit{fewest pending messages}, which are closest to graduation and least likely to be reloaded repeatedly from disk.

\ta's eviction policy maintains an ordering of \IB vertices by pending message count while supporting insertions, removals, score updates, and selection of the $k$ lowest-scoring vertices. Rather than a conventional Python heap, it exploits the bounded integer score range $[1,\texttt{max\_in\_degree}]$, where a vertex's score equals its pending message count~($0$ means \CP). Vertices are organized into score-indexed buckets implemented as doubly linked lists, enabling $\mathcal{O}(1)$ insertion, removal, and score updates. Eviction proceeds by scanning the lowest non-empty buckets, yielding $\mathcal{O}(k)$ selection of $k$ eviction candidates.

\subsubsection{Cold Store} 
The cold store is a NumPy \texttt{mmap} file. Unlike single-pass feature and embedding scans, it uses buffered I/O because evicted states may be reloaded in later chunks, allowing the OS page cache to absorb eviction--reload traffic.

\subsection{Graduation Processor}\label{subsec:grad-processor}\label{subsec:sys-grad}
The \textit{graduation processor} handles completed vertices~(Fig.~\ref{fig:atlas-arch}, \textit{blue}). 
When a vertex's pending message count reaches zero, the \textit{orchestrator} instructs the \textit{memory manager} to finalize its aggregation, 
append it to a configurable \textit{graduation buffer}, and release the hot-store slot.

Once full, a graduation buffer is enqueued for GPU offload. Double buffering lets one buffer collect newly completed vertices while the other is processed asynchronously. A dedicated GPU-offload thread applies the layer transformation using CUDA streams to overlap transfers and compute, preventing the orchestrator from blocking. Transformed embeddings are then enqueued to the \textit{write queue}.

\subsection{Embedding Writer}\label{subsec:sys-writer}
Transformed embeddings are materialized on disk for the next layer or final output. A dedicated writer consumes embeddings from the \textit{write queue} and stages them in GPU-resident partition buffers. Since vertices arrive in graduation rather than vertex-ID order, the writer range-partitions them by vertex ID; when a partition buffer fills, it is sorted on the GPU and flushed sequentially as a sorted \textit{run file}.

\ta writes run files through \texttt{kvikio}/cuFile, 
using GPUDirect Storage (GDS) when supported by the platform, and cuFile compatibility mode otherwise.
On platforms where true GDS is unavailable, such as our consumer-GPU machine, we use cuFile-managed \textit{bounce buffers\footnote{https://docs.nvidia.com/gpudirect-storage/api-reference-guide/\#cufile-compatibility-mode}}, but the interface remains unchanged. Even without true GPU-to-storage DMA, GPU-resident partition buffers avoid host-side sorting, reduce RAM pressure and leave more memory for the hot store.

\subsection{Topology-aware Graph Reordering}\label{subsec:sys-order}
\begin{figure}[t]
    \centering
    \subfloat[Inference time vs.\ $C(\pi)$\label{subfig:span-time}]{
        \includegraphics[width=0.34\linewidth]{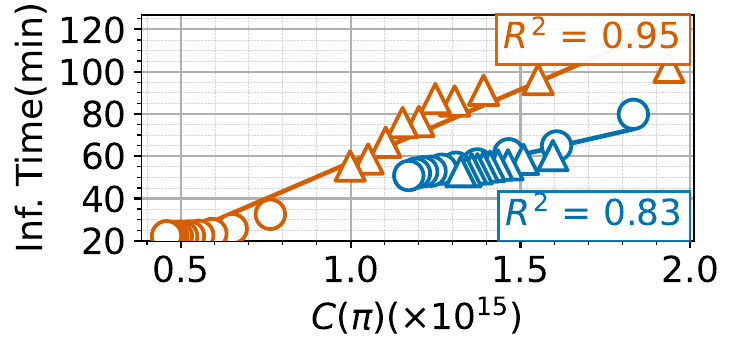}
    }
    \subfloat[Cold reloads vs.\ $C(\pi)$\label{subfig:span-reloads}]{
        \includegraphics[width=0.34\linewidth]{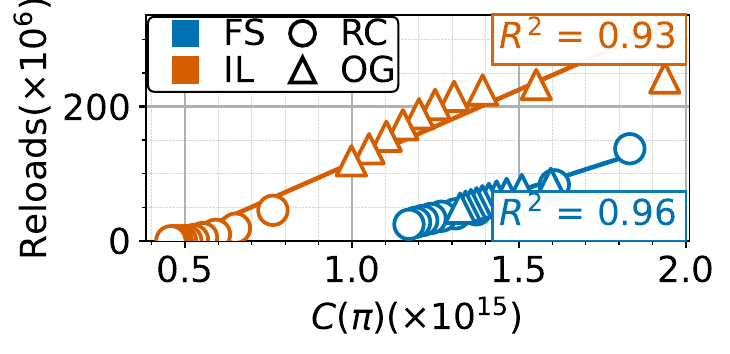}
    }\\
    \subfloat[$J$ vs.\ $C(\pi)$\label{subfig:j_vs_c}]{
        \includegraphics[width=0.34\linewidth]{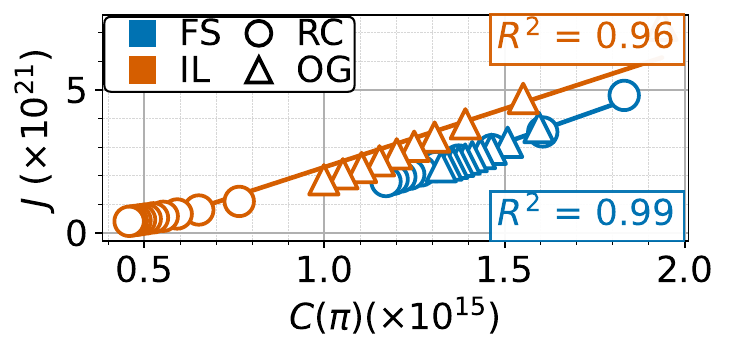}
    }
    \caption{Topology-based reordering on Friendster~(FS) and IGB-large~(IL) with RCMK~(RC) and Original~(OG) bootstraps, showing correlation between total span $C(\pi)$ and (a) total inference time; (b) number of reloads from cold store, i.e., I/O cost; and (c) our proposed objective $J$. Each point corresponds to an intermediate ordering generated during Alg.~\ref{alg:ta} starting from OG and RC bootstraps.}
    \label{fig:span}
\end{figure}

Processing vertices in the original vertex-ID order can significantly increase hot-store residency. Vertices processed early may not complete until much later, leaving their partial states resident for extended periods. This increases memory pressure, eviction frequency, and ultimately execution time. While the \at greedy reordering strategy maximizes completion rate, it does not explicitly minimize the \emph{span}, i.e., the interval between a destination's first and last incoming message. Next, we characterize vertex residency and I/O overhead in terms of vertex span and provide empirical evidence in Fig.~\ref{fig:span}.

Let $G=(V,E)$ be a directed graph, where $\pi(v)$ denotes the processing rank of vertex $v$, and $N^{-}(v)$ its in-neighbors. A vertex becomes \textit{active} when it receives a message from its first in-neighbor and \textit{completes} after receiving the message from its last. Accordingly, its activation and completion ranks are $a(v)=\min_{u\in N^{-}(v)}\pi(u)$ and $c(v)=\max_{u\in N^{-}(v)}\pi(u)$, respectively, yielding the \textit{span} $L(v)=c(v)-a(v)$. 
Let $A(t)$ denote the number of active vertices at processing rank $t$. Then:
\begin{align*}
    A(t)=\sum_{v\in V}\mathbf{1}\bigl[a(v)\le t < c(v)\bigr],    
\end{align*}
and the cumulative active-state occupancy is
\begin{align*}
    M=\sum_t A(t)=\sum_t \sum_{v\in V}\mathbf{1}\bigl[a(v)\le t < c(v)\bigr] \\ = \sum_{v\in V} \sum_t \mathbf{1}\bigl[a(v)\le t < c(v)\bigr] = \sum_{v\in V} L(v) = C(\pi)
\end{align*}

\textit{Thus, reducing $C(\pi)$ lowers cumulative residency, eviction pressure and I/O.} The objective is:
\begin{align*}
    \min_\pi C(\pi)
    = \min_\pi \sum_{v\in V}\Bigl(\max_{u \in N^{-}(v)} \pi(u) - \min_{u \in N^{-}(v)} \pi(u)\Bigr).
\end{align*}

Bandwidth-minimization heuristics such as RCMK~\cite{chan1980linear} optimize $B=\max_{(u,v)\in E}|\pi(u)-\pi(v)|$, a worst-case edge-length metric. Since all in-neighbors of $v$ lie within $[\pi(v)-B,\pi(v)+B]$, $L(v)\le 2B$ and $C(\pi)\le 2|V|B$. This loose bound makes RCMK a useful bootstrap, but it does not directly minimize total span.

\begin{algorithm}[t]
\caption{\ta~topology-aware reordering}
\label{alg:ta}
\footnotesize
\begin{algorithmic}[1]
\Require Graph $G=(V,E)$, initial ordering $\pi$
\State Initialize $r(v)\gets\pi(v)$
\Repeat
    \State $\mu(v)\gets\frac1{d^-(v)}\sum_{u\in N^-(v)}r(u),\ \forall v$
    \State $r(w)\gets
    \dfrac{\sum_{v\in N^+(w)}\mu(v)/d^-(v)}
          {\sum_{v\in N^+(w)}1/d^-(v)},\ \forall w$
    \State $\pi\gets\textsc{ArgSort}(r)$
\Until{$\pi$ converges}
\State \Return $\pi$
\end{algorithmic}
\end{algorithm}

Since optimizing $C(\pi)$ over $|V|!$ orderings is infeasible and bandwidth only bounds it loosely, \ta instead \textit{minimizes in-neighborhood dispersion}. Each vertex receives a real-valued position $r(v)$, and we measure the spread of the in-neighbors of each destination around a center $\mu(v)$:
\begin{align*}
    J(r, \mu) = \sum_{v \in V} \frac{1}{d^-(v)} \sum_{u \in N^-(v)} \bigl(r(u) - \mu(v)\bigr)^2.
\end{align*}
The $1/d^-(v)$ normalization factor gives each destination equal weight, preventing high-degree vertices from dominating the objective. In GNNs, each vertex aggregates a self message, so $d^-(v)\ge 1$ (normalization and $\mu(v)$ are well-defined); a vertex with only its self message has $L(v)=0$.

Since $J$ is differentiable, we iteratively update the locality centers $\mu$ and the vertex positions $r$. For a fixed ordering $r$, setting $\partial J/\partial\mu(v)=0$ gives
\begin{align*}
    \mu^*(v) = \frac{1}{d^-(v)} \sum_{u \in N^-(v)} r(u),
\end{align*}
the average position of a vertex's in-neighbors. Likewise, for fixed locality centers $\mu$, setting $\partial J/\partial r(w)=0$ gives
\begin{align*}
    r^*(w) = \frac{\displaystyle\sum_{v \in N^+(w)} \frac{\mu(v)}{d^-(v)}}{\displaystyle\sum_{v \in N^+(w)} \frac{1}{d^-(v)}},
\end{align*}
the weighted average of the locality centers of its out-neighbors~($N^+(w)$). Sorting the updated positions yields the new ordering. These update rules naturally give rise to the iterative reordering procedure shown in Alg.~\ref{alg:ta}.

Fig.~\ref{fig:span} validates this on FS and IL (Tab.~\ref{tab:datasets}) using OG and RCMK initializations; each point is an intermediate ordering from Alg.~\ref{alg:ta}.
Figs.~\ref{subfig:span-time} and~\ref{subfig:span-reloads} show that $C(\pi)$ strongly predicts inference time and cold reloads~($R^2{=}0.83$--$0.96$), while Fig.~\ref{subfig:j_vs_c} shows that the neighborhood dispersion objective $J$ closely tracks $C(\pi)$~($R^2{=}0.96$--$0.99$). 
Thus, reducing $J$ lowers span, I/O, and runtime.

\subsection{Generalizability of \ta}\label{subsec:sys-generalize}
\ta extends beyond GCN/GIN-style aggregation to other message-passing GNNs: SAGEConv and GATConv, and to sampled inference.

\paragraph{SAGEConv}\label{para:generalize-sage}
SAGEConv can double hot-store footprint because its update concatenates a vertex's aggregated neighborhood representation with its own embedding before applying a shared transformation:
\begin{align*}
    h_v^{(\ell+1)}=\sigma\left(W[h_{\mathrm{agg}, v}\|h_v^{(\ell)}]\right), \text{where} \\
    h_{\mathrm{agg}, v}=\textsc{Aggregate}_{u\in N^-(v)}h_u^{(\ell)}
\end{align*}
Under broadcast execution, $h_v^{(\ell)}$ is immediately available when $v$ is streamed, whereas $h_{\mathrm{agg},v}$ is ready only after all propagated messages arrive; retaining both vectors doubles per-vertex state (as seen in \at~\cite{atlas}).
\ta eliminates this overhead by partitioning the neural-network matrix $W \in \mathbb{R}^{d'\times2d}$ into $W=[W_{\mathrm{agg}}~W_{\mathrm{self}}]$, where $W_{\mathrm{agg}},W_{\mathrm{self}}\in\mathbb{R}^{d'\times d}$. 
Therefore, by linearity, $W[h_{\mathrm{agg}}\|h_v^{(\ell)}]=W_{\mathrm{agg}}h_{\mathrm{agg}}+W_{\mathrm{self}}h_v^{(\ell)}$, which is exactly equivalent to the original formulation. 
Thus, the hot store maintains only $h_{\mathrm{agg}}$; once complete, \ta writes $W_{\mathrm{agg}}h_{\mathrm{agg}}$ as a run file. A subsequent sequential pass computes $W_{\mathrm{self}}h_v^{(\ell)}$, sums the projected terms, applies bias/activation if present and emits the final embedding. This halves hot-store state and reduces eviction--reload cycles while preserving SAGEConv semantics, at the cost of one extra sequential pass.

\paragraph{GATConv}\label{para:generalize-gat}
GATConv is more challenging because attention depends on both source and destination embeddings and is edge-specific:
\begin{align*}
    h_v^{(\ell+1)}&=\sigma\left(\sum_{u\in\mathcal N^-(v)}\alpha_{uv}Wh_u\right), \quad \text{where}
    \\
    \alpha_{uv}&=\mathrm{softmax}(e_{uv}), \quad \text{and}
    \\
    e_{uv}&=\mathrm{LeakyReLU}\left(\mathbf a^\top[Wh_u\|Wh_v]\right)
\end{align*}
\ta supports GATConv using multiple sequential passes, avoiding $\mathcal{O}(|E|d')$ edge-feature materialization. First, it materializes transformed features $h'_v=Wh_v$ for all vertices, reducing later passes from dimension $d$ to $d'$.
This is followed by a topology-only pass over the CSR that computes edge attention scores and applies a neighborhood-wise softmax to obtain the attention coefficients $\alpha_{uv}$. 
Finally, \ta streams the transformed embeddings again and aggregates using these attention weights. This preserves GAT semantics while trading random gathers and high-dimensional edge state for additional sequential passes.
For multi-head GAT, the same procedure applies per head, with proportional storage and pass costs.

\paragraph{Sampling-based Inference}\label{para:generalize-sample}
Some applications require deterministic full-neighborhood inference, while others tolerate fanout-sampled approximation. \ta supports both.
For fanout $k$, \ta constructs a layer-specific edge mask by sampling up to $k$ incoming edges per vertex and suppressing the rest.
Sampling reduces propagated messages, but not the sequential scan of vertex features/embeddings needed to compute updated representations.
Thus, \ta retains the same sequential read pattern while broadcasting only over sampled edges, reducing hot-store residency, eviction traffic and runtime.


\begin{table}[t]
    \centering
    \caption{Graph datasets~\cite{ogb, igb, leskovec2020snap} used in experiments.}
    \label{tab:datasets}
    \setlength{\tabcolsep}{1pt}
    \footnotesize
    \begin{tabular}{l||c|c|c|c|c}
        \hline
        \textbf{} & \textbf{\textit{Papers}} & \textbf{\textit{Friendster}} & \textbf{\textit{MAG-Cites}} & \textbf{\textit{IGB-Large}} & \textbf{\textit{IGB-Full}} \\ \hline \hline
        \textbf{\textit{Abbr.}} & PA & FS & MA & IL & IF \\ 
        \textbf{\textit{\# Vertices}} & $111\M$ & $65\M$ & $121\M$ & $100\M$ & $269\M$\\ 
        \textbf{\textit{\# Edges}} & $1.7\B$ & $3.6\B$ & $1.4\B$ & $1.2\B$ & $4\B$  \\ 
        \textbf{\textit{Feat. Dim}} & $128$ & $1024$ & $768$ & $1024$ & $1024$ \\ 
        \textbf{\textit{\# Classes}} & $172$ & $64$ & $153$ & $19$ & $19$ \\ 
        \textbf{\textit{Top. Size~(GiB)}} & $14$ & $28$ & $12$ & $10$ & $32$ \\ 
        \textbf{\textit{Feat. Size~(GiB)}} & $54$ & $251$ & $350$ & $382$ & $514$~(FP$16$) \\ \hline
    \end{tabular}
\end{table}

\section{Evaluation}\label{sec:eval}
We evaluate \ta along four axes: end-to-end performance against OOC baselines, support for exact and sampled inference, impact of runtime components and resource usage.

\subsection{Experimental Setup}\label{subsec:eval-setup}
We evaluate \ta using four 2-layer vertex-classification GNNs: GraphConv~(\textbf{GCN})~\cite{kipf2017gcn}, SAGEConv~(\textbf{SAGE})~\cite{hamilton2017sage}, GINConv~(\textbf{GIN})~\cite{gin}, and single-head GATConv~(\textbf{GAT})~\cite{gat}, all with hidden dimension $128$.
Experiments use five open-source citation and social-network datasets~(Tab.~\ref{tab:datasets}), with feature stores ranging from $54$~GiB for \textit{Papers}~\cite{ogb} to $514$~GiB for \textit{IGB-Full}~\cite{igb}. While IGB-Full uses FP16, rest use FP32 precision.
We report both exact full-neighborhood inference and fanout-sampled inference.
For exact full-neighborhood inference, \ta matches the output of an in-memory layer-wise DGL~\cite{wang2019deep} implementation on PA using identical weights and precision, with mean per-vertex max absolute error $8\times10^{-5}$ and mean relative error $2.8\times10^{-6}$.

Unless otherwise stated, experiments run on a single workstation with a $12$-core AMD Ryzen $9$ $9900$X CPU~($4.4$\,GHz), $128$\,GiB RAM, an NVIDIA RTX $5090$ GPU with $32$\,GiB VRAM, a $2$\,TiB Samsung $990$ PRO SSD, and Ubuntu $24.04.3$ LTS. We clear the OS page cache before each run.

\begin{figure*}[t]
    \centering
    \includegraphics[width=\textwidth]{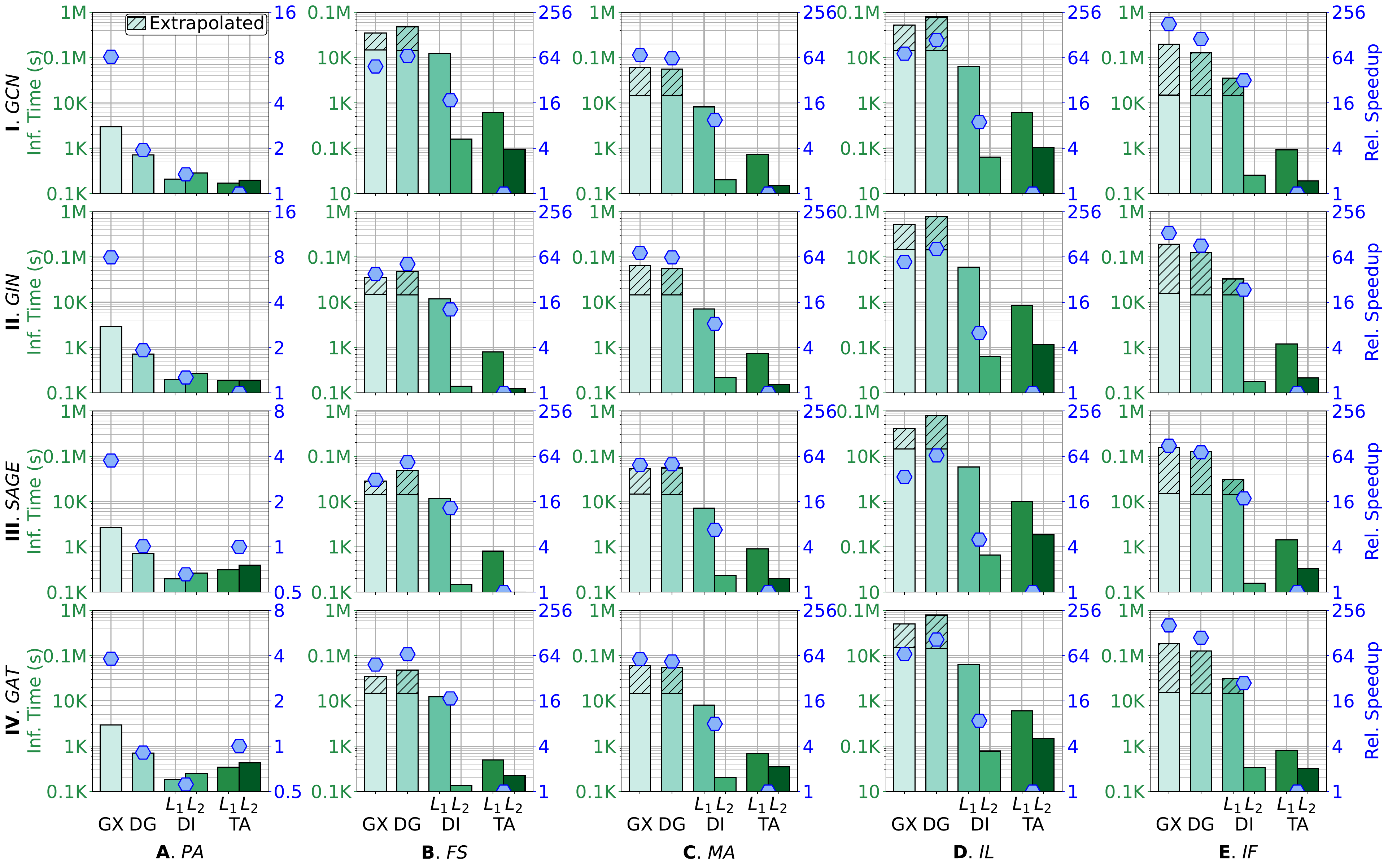}
    \caption{Sampled inference performance~(\textit{fanout=10} per layer) of \textsf{\underline{TA}urus} vs. \underline{G}ine\underline{X}, \underline{DG}L, and \underline{D}G\underline{I} for 2-layer GCN, SAGE, GIN and GAT on PA, FS, MA, IL and IF. Bars show total inference time~(s, log scale); markers show speedup over \ta. Hatched bars denote extrapolated runtimes.}
    \label{fig:overview-plots}
\end{figure*}

\subsection{\ta Implementation and Baselines}
\ta is implemented in Python with NumPy v2.0 and PyTorch v2.8; the graph reader and embedding writer are C++ PyTorch extensions compiled with \texttt{ninja}. Unless varied in ablations, we use 8\,MiB chunks, 256\,MiB graduation buffers, queue size 20, 50 GiB hot store for PA/MA/IL, 100\,GiB for FS/IF, and a 16\,GiB GPU store after reserving VRAM for write buffers, intermediate tensors, and CUDA context. Framework overhead is 5--7\,GiB.

We compare \ta~(\textbf{TA}) with three OOC baselines: vertex-wise Ginex~\cite{ginex}~(\textbf{GX}) and DGL GraphBolt OnDisk\footnote{https://www.dgl.ai/dgl\_docs/generated/dgl.graphbolt.OnDiskDataset.html}~(\textbf{DG}), and layer-wise DGI~\cite{dgi}~(\textbf{DI}).

Ginex targets OOC GNN training with disk-resident neighbor and feature caches. We adapt it to inference by disabling backpropagation and using its default superbatch/batch sizes~(2500--3300/1000), with 90\,GiB feature cache and 10\,GiB neighbor cache on our 128\,GiB system.
DGL GraphBolt OnDisk uses DGL's \texttt{OnDiskDataset} abstraction for vertex-wise OOC execution. We use RCMK reordering~\cite{chan1980linear} and batch size 32K to improve locality, retaining other defaults.
DGI is a layer-wise inference framework using dynamic batching, RCMK ordering~\cite{chan1980linear}, and NumPy \texttt{mmap} files for features and CSC indices. We use the paper's default settings.

We report layer-wise results for DI and TA, and end-to-end results for GX and DG. 
Runs are capped at $4$\,h; incomplete runs are linearly extrapolated from completed vertex ranges/chunks, per layer for DI/TA and end-to-end for GX/DG. If a DI layer times out, later layers are measured with dummy inputs of matching dimensions and added to the extrapolated incomplete layer.

\subsection{Comparison with Baselines}
Fig.~\ref{fig:overview-plots} compares sampled inference~(\textit{fanout=10}) across four 2-layer GNNs and five datasets. We use fanout 10 because smaller fanouts understate message-propagation costs, while larger fanouts make most baselines exceed the time budget. 
GX/DG bars are end-to-end times; DI/TA report per-layer times, and the total inference time is the sum across layers.

\paragraph{Performance Improvements over Vertex-Wise Baselines} 
TA outperforms GX by $\approx40\times$, $62\times$, $57\times$ and $140\times$ on OOC FS, MA, IL and IF, respectively, because 
GX incurs repeated cache construction, sampled-batch materialization and feature I/O; Belady's policy reduces but does not eliminate the latter.
TA similarly outperforms DG by $\approx60\times$, $57\times$, $90\times$ and $96\times$, as DG gathers disk-resident features per batch.
In contrast, TA streams embeddings sequentially: GCN/GIN require one read/write pass per layer, while SAGE/GAT add sequential passes but avoid repeated random gathers. Accordingly, DG and GX read up to $\approx50\times$ and $\approx108\times$ more data, respectively, in the evaluated workloads.

\paragraph{Performance Improvements over Layer-Wise DGI} 
Combined across both layers, TA achieves average speedups of $15\times$, $8.2\times$, and $7.2\times$ over DI on FS, MA, and IL, respectively (Fig.~\ref{fig:overview-plots}, cols. B--D). 
On IF, DI's first layer exceeded the $4$\,h cap; we estimate total DI time by extrapolating that layer and measuring later layers with matching dummy inputs.
TA outperforms DI by $\approx25\times$ averaged across all models, while completing inference on the $514$ GiB dataset in under $30$ min. 
DI remains stronger than vertex-wise baselines because layer-wise execution avoids redundant neighborhood expansion, but still reads up to $10\times$ more data than TA.

\paragraph{Performance When Features Fit in Memory}
On PA~(Fig.~\ref{fig:overview-plots}, col. A), the graph topology and features fit entirely in memory, allowing the OS page cache to eliminate most disk I/O. Even so, TA outperforms the vertex-wise GX baseline by $3.7$--$8.1\times$, and DG by $\approx1.9\times$ for GCN and GIN, because it avoids repeated reads. Compared to the layer-wise DI baseline, TA remains up to $\approx1.3\times$ faster for GCN and GIN, but is $\approx0.6\times$ slower for SAGE and GAT. Since I/O costs are mitigated by the OS page cache, the additional passes in TA to support SAGE and GAT dominate execution, while DI executes these operators directly in memory.

\paragraph{Reduction in Layer Execution Time}
Across the FS, MA, IL, and IF datasets, \ta's execution time drops sharply from $L_1$ to $L_2$ by $\approx4.8\times$ across all models, because the first layer projects high-dimensional inputs~($768/1024$) to a $128$-dimensional hidden space. Consequently, subsequent layers stream significantly less data from disk, reducing both I/O overhead and computational costs due to reduced dimensionality. PA exhibits the opposite trend because its input and hidden dimensions are $128$, while the output dimension is $172$. Hence, $L_2$ is computationally costlier than $L_1$, resulting in an average $\approx17\%$ increase in execution time across all models.

\paragraph{Single vs. Multi Pass GNNs}
SAGE and GAT require multiple sequential passes, increasing runtime relative to GCN/GIN.
Compared to GCN, SAGE/GAT are $93\%$/$112\%$ slower on PA and $24$--$64\%$/$0.7$--$17\%$ slower on OOC datasets, consistent with reading $103\%$/$51\%$ more data.
The increase is largest on PA because the graph and its features largely fit in memory; hence, the layer decomposition and repeated scans become costlier than out-of-the-box GNN operators.
Lastly, the relatively small increase in GAT execution time compared to SAGE, despite the additional passes, is expected because GAT first projects the embeddings from $d$ to $d'$~($d' < d$), allowing all subsequent passes to operate on the lower-dimensional representation.~(\S~\ref{para:generalize-gat}).

\subsection{Impact of \ta Ordering}

\begin{figure}[t]%
    \centering%
    \subfloat[FS/GCN2\label{subfig:span-fs}]{%
        \includegraphics[width=0.35\columnwidth]{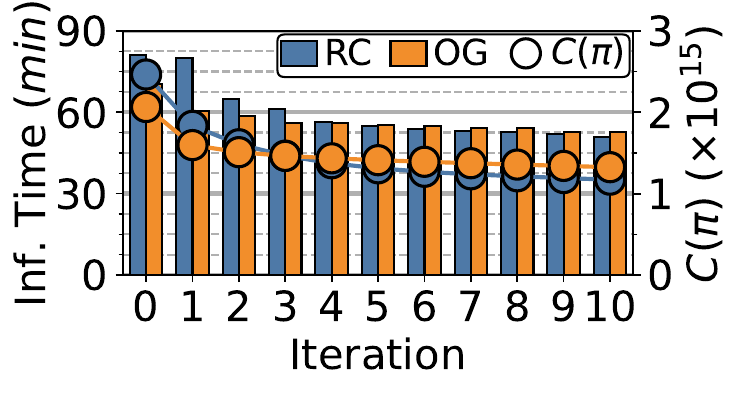}%
    }\qquad
    \subfloat[IL/GCN2\label{subfig:span-il}]{%
        \includegraphics[width=0.35\columnwidth]{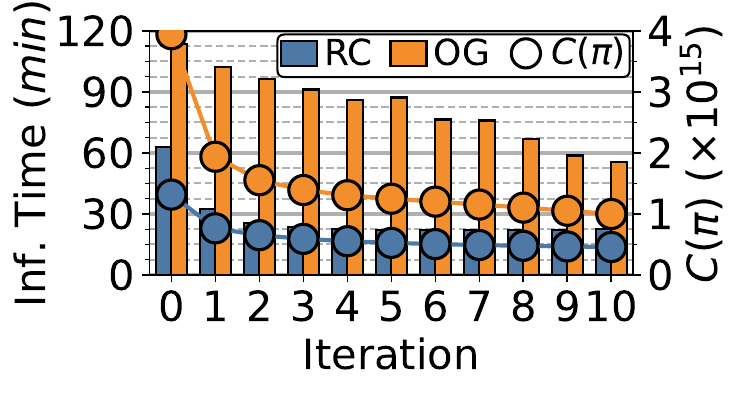}%
    }
    \caption{Impact of \ta reordering bootstrap and iterations showing end-to-end inference time~(\textit{bars}, left Y axis) and span cost~$C(\pi)$~(\textit{markers}, right Y axis). RC and OG mean RCMK and original ordering bootstraps, respectively.}
    \label{fig:reorder-1}
\end{figure}

\begin{figure}[t]
    \centering
    \subfloat[Inf. time, I/O time, and \# reloads.\label{subfig:reorder-2.1}]{
        \includegraphics[width=0.32\columnwidth]{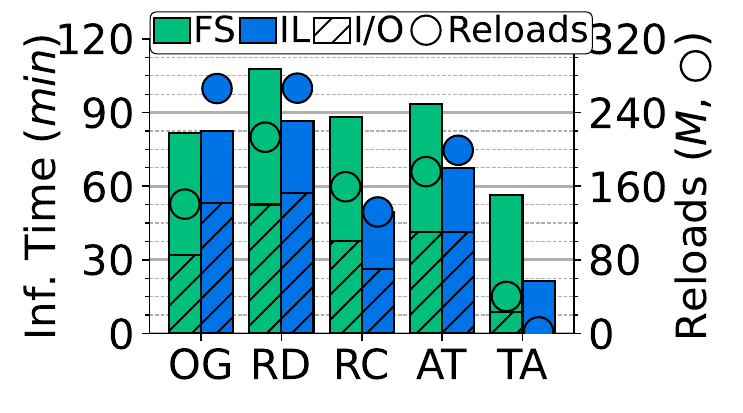}
    }\qquad
    \subfloat[Dst. distribution and hot-store reuse.\label{subfig:reorder-2.2}]{
        \includegraphics[width=0.4\columnwidth]{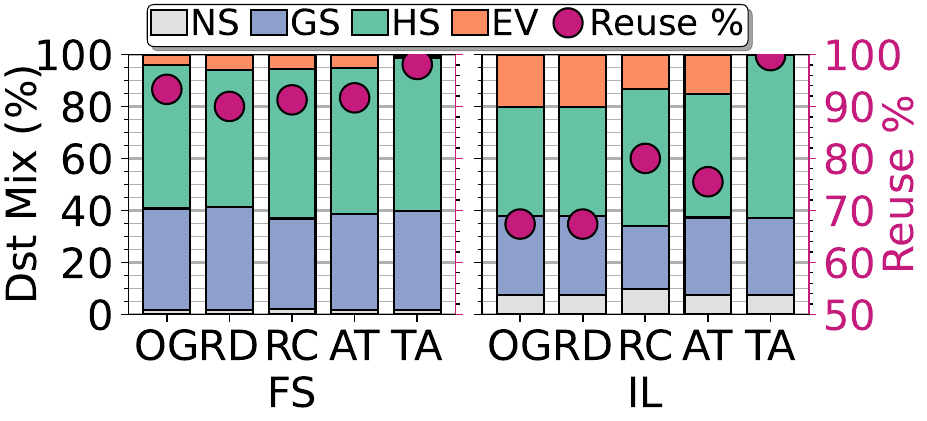}
    }
    \caption{Impact of \ta reordering on execution behavior. (a) E2E inference time~(\textit{bars}, left Y axis), I/O time~(\textit{hatched bars}, left Y axis), and \# reloads~(\textit{markers}, right Y axis); (b) Avg. destination-state split~(\textit{stacked bars}, left Y axis) and hot-store reuse~(\textit{markers}, right Y axis) for Layer 1.}
    \label{fig:reorder-2}
\end{figure}

We compare OG, RD, RCMK~(RC), \at~(AT) and \ta~(TA) orderings for exact 2-layer GCN inference on IL and FS, using $50$/$80$~GiB hot stores, a $10$~GiB GPU store and TA eviction~(Fig.~\ref{fig:reorder-2}). We also evaluate convergence under OG and RC bootstraps~(Fig.~\ref{fig:reorder-1}).

\paragraph*{Convergence of \ta reordering} Fig.~\ref{fig:reorder-1} shows the span objective $C(\pi)$~(\S~\ref{subsec:sys-order}) decreases monotonically over $10$ iterations across both OG and RC bootstraps for both datasets. This confirms that the iterative TA updates consistently improve the ordering.
On FS~(Fig.~\ref{subfig:span-fs}), $C(\pi)$ decreases by $36$--$52\%$~(\textit{markers}, right Y axis), translating to a $25$--$38\%$ reduction in end-to-end inference time~(\textit{bars}, left Y axis).
The gains are larger on IL~(Fig.~\ref{subfig:span-il}), where $C(\pi)$ decreases by $65$--$75\%$, yielding a $51$--$64\%$ reduction in runtime. 
RC is the stronger bootstrap, giving $6$--$60\%$ lower final runtimes after convergence~($\approx50$ vs. $53$ min for FS; $\approx22$ vs. $55$ min for IL), so \ta uses RC by default.
Lastly, over $84$--$90\%$ of the total span reduction is achieved within the first $4$--$5$ iterations, indicating rapid convergence. Based on this observation, we limit the number of iterations to $5$ in \ta.

\paragraph*{Impact on System Performance} Fig.~\ref{subfig:reorder-2.1} compares the impact of different vertex orderings on system performance. 
TA uses RC as bootstrap with $5$ iterations of refinement. 
By reducing span, TA consistently exhibits the lowest end-to-end inference time, reducing runtime by $\approx31$--$48\%$ over OG, RD, AT, and RC on FS to $56$ min~(\textit{green bars}, left Y axis) and by $\approx57$--$75\%$ on IL to $21$ min~(\textit{blue bars}, left Y axis). The improved span substantially lowers SSD traffic, reducing I/O time~(both reads/writes) by $3.6$--$5.9\times$ on FS~(\textit{hatched green bars}, left Y axis) and $73$--$158\times$ on IL~(\textit{hatched blue bars}, left Y axis), while also decreasing eviction-reload cycles by $4.3\times$ and $126\times$, respectively, on average~(\textit{markers}, right Y axis). 
On IL, TA reduces I/O overhead to near-negligible levels under the same GPU and hot-store budgets as the other orderings.

\paragraph*{Hot Store Utilization and Reuse} Finally, we quantify this benefit at the chunk level in Fig.~\ref{subfig:reorder-2.2}, showing the average split of destination vertices for a chunk into HS~(\IB), GS~(\OG), NS~(\NS), and EV~(\EV) states. TA increases the mean \% of destination vertices in the hot store~(HS) by $\approx3.7\%$ and $17\%$~(\textit{green stack}, left Y axis), with a corresponding decrease in the \% in the cold store~(EV) by $\approx4\%$ and $17\%$~(\textit{orange stack}, left Y axis) for FS and IL, respectively. Consequently, the hot-store reuse, i.e., the fraction of \textit{active} destinations~(vertices with $\ge 1$ messages received) already resident in the hot store~($\frac{HS}{HS+EV}$), increases from $\approx90$--$93\%$ to $98\%$ on FS and from $\approx67\%$ to $99.8\%$ on IL~(\textit{markers}, right Y axis). 
By keeping nearly all active states resident until graduation, TA reduces I/O thrashing.

\subsection{Impact of \ta Eviction Policy}

\begin{figure}[t]
    \centering
    \subfloat[Inf. time, I/O time, and \# reloads.\label{subfig:eviction-1}]{
        \includegraphics[width=0.34\columnwidth]{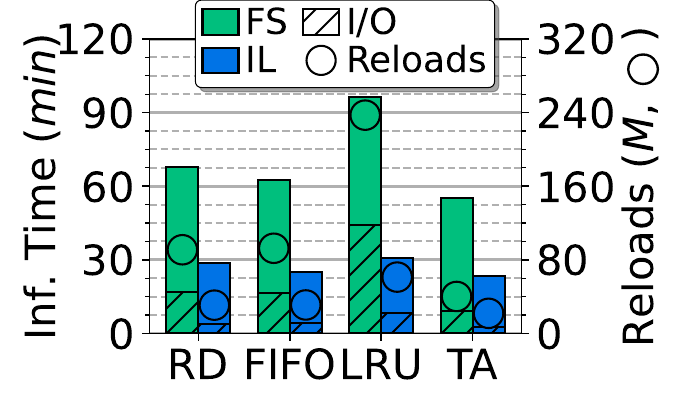}
    }\qquad\qquad
    \subfloat[CDF of \# reloads.\label{subfig:eviction-2}]{
        \includegraphics[width=0.44\columnwidth]{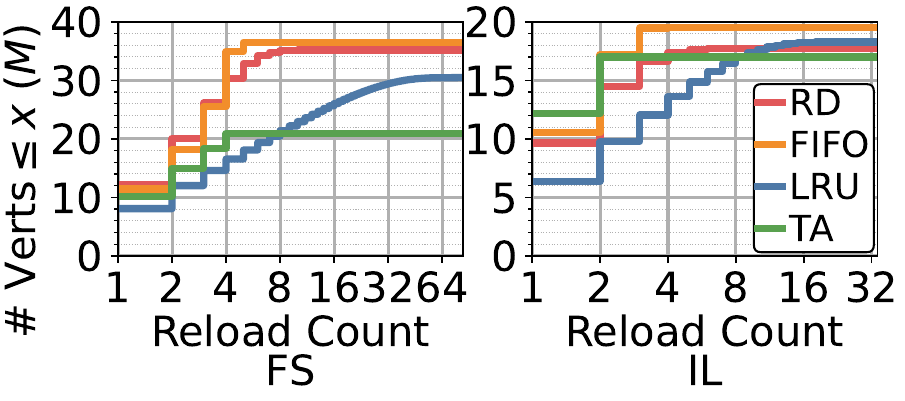}
    }
    \caption{Impact of \ta eviction on execution behavior. (a) E2E inference time~(\textit{solid bars}, left Y axis), I/O time~(\textit{hatched bars}, left Y axis), and \# reloads~(\textit{markers}, right Y axis); (b) Cumulative reload distribution of the number of unique vertices reloaded from the cold store.}
    \label{fig:eviction}
\end{figure}

Fig.~\ref{fig:eviction} compares the proposed minimum pending-messages eviction policy~(TA) against random~(RD), first-in-first-out~(FIFO), and least-recently-used~(LRU) eviction for exact 2-layer GCN inference under TA ordering, using $80$~GiB and $40$~GiB hot stores for FS and IL, respectively, and a $10$~GiB GPU store.

Fig.~\ref{subfig:eviction-1} shows that prioritizing vertices for eviction with the fewest pending messages substantially improves execution time. Compared to RD and LRU, TA reduces inference time by $\approx19$--$43\%$ on FS and $\approx20$--$26\%$ on IL (\textit{solid bars}, left Y axis). 
LRU performs poorly because it repeatedly evicts vertices far from completion, while FIFO improves upon LRU by providing a minimum residency period after admission. 
TA outperforms both by evicting vertices closest to completion, allowing them to graduate soon after reload and minimizing eviction--reload cycles. Consequently, TA reduces I/O time by $\approx31$--$80\%$ to $9$ min and $2.7$ min (\textit{hatched bars}, left Y axis), while reducing hot-store reloads by $\approx1.4$--$6\times$ to $40\M$ and $21\M$ on FS and IL, respectively (\textit{markers}, right Y axis).

This anti-thrashing behavior is directly validated in Fig.~\ref{subfig:eviction-2} showing the distribution of destination reload counts. 
Because TA selects near-complete victims, most vertices see only $2$--$4$ reloads across FS and IL, versus $12$--$17$ for RD. LRU has a long tail of $34$--$84$ reloads because inactive but far-from-complete vertices are repeatedly swapped.
By intelligently selecting victims that will rapidly graduate upon reload, TA minimizes repeated cold-store accesses, drastically reducing overall execution time.

\subsection{Impact of Hot Store Size}

\begin{figure}[t]
    \centering
    \subfloat[FS/GCN2\label{subfig:hs-1}]{
        \includegraphics[width=0.27\columnwidth]{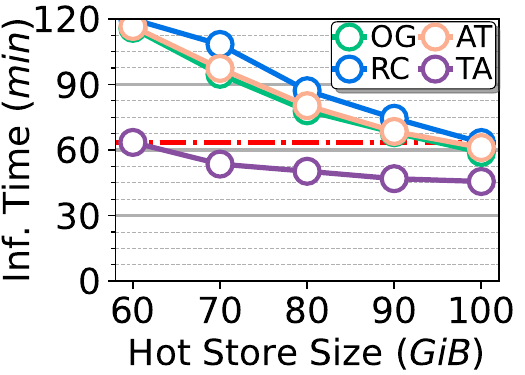}
    }\qquad\qquad
    \subfloat[IL/GCN2\label{subfig:hs-2}]{
        \includegraphics[width=0.27\columnwidth]{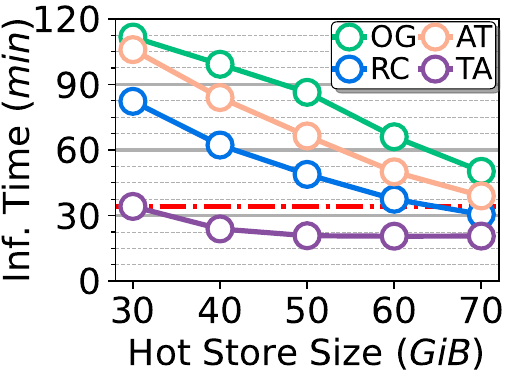}
    }
    \caption{Impact of hot-store capacity on E2E inference time under different vertex orderings for (a) FS and (b) IL. TA achieves identical performance with substantially lower hot-store budgets than OG, RC, and AT~(\textit{dashed red line}).}
    \label{fig:hotstore-reorder}
\end{figure}

\begin{figure}[t]
    \centering
    \subfloat[FS/TA/GCN2\label{subfig:hs-3}]{
        \includegraphics[width=0.28\columnwidth]{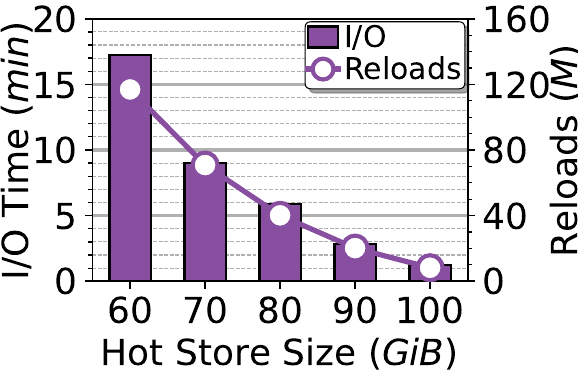}
    }\qquad\qquad
    \subfloat[IL/TA/GCN2\label{subfig:hs-4}]{
        \includegraphics[width=0.28\columnwidth]{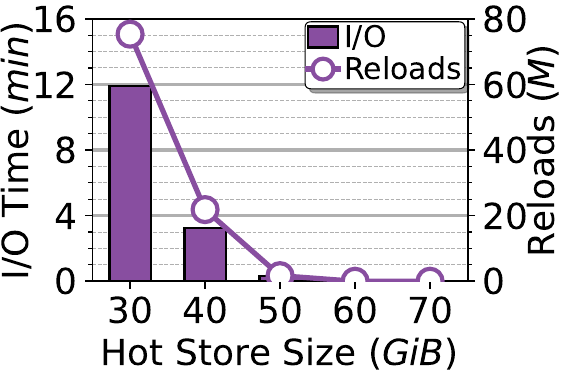}
    }
    \caption{Impact of hot-store capacity on TA. SSD I/O time~(\textit{bars}, left Y axis) and cold-store reloads~(\textit{markers}, right Y axis) under TA reorder for (a) FS and (b) IL.}
    \label{fig:hotstore}
\end{figure}

Figs.~\ref{fig:hotstore-reorder} and~\ref{fig:hotstore} study the impact of varying hot-store sizes for exact 2-layer GCN inference on FS and IL, respectively, using a fixed $10$~GiB GPU store.

\textit{Larger hot stores only partially compensate for poor ordering.} Figs.~\ref{subfig:hs-1} and~\ref{subfig:hs-2} show that increasing hot-store capacity reduces inference times by allowing more partially aggregated vertices to remain resident, thereby relieving eviction pressure, across all reorder strategies. However, the steep performance curves for OG, RC, and AT indicate that these orderings are fundamentally memory-starved. In contrast, TA exhibits a notably flatter curve, showing that it is consistently the least sensitive to hot-store capacity. While increasing the RAM budget improves TA by only $\approx28$--$40\%$, the baseline orderings see improvements of $\approx50$--$65\%$ across FS and IL. Consequently, TA effectively decouples performance from memory scale, requiring substantially less RAM to achieve identical throughput. For example, on FS, TA at just $60$~GiB matches the peak performance of AT/OG at $\approx90$~GiB. Similarly, on IL, TA at $30$~GiB matches RC at $\approx60$~GiB~(\textit{dashed red line}). By minimizing the average lifespan of active vertices~(\S~\ref{subsec:sys-order}), TA ensures that partial states graduate quickly, drastically reducing the structural need for massive buffer capacity.

\textit{TA approaches peak throughput with modest memory.}
Figs.~\ref{fig:hotstore-reorder} and \ref{fig:hotstore} show that the $28$--$40\%$ reduction in end-to-end inference time achieved by TA is primarily due to lower SSD I/O overhead. As the hot-store size increases, reloads decrease from $\approx120\M$ to $10\M$ on FS and from $\approx75\M$ to nearly zero on IL~(\textit{circles}, right Y axis), reducing I/O time from $\approx17$ min to $2$ min on FS and from $\approx12$ min to negligible on IL~(\textit{bars}, left Y axis). For IL, reloads are almost completely eliminated beyond a $50$~GiB hot store. At this point, the hot store can hold nearly all active vertex states, eliminating eviction--reload cycles. Consequently, SSD I/O becomes negligible, inference becomes compute-bound, and further increasing the hot-store size provides little additional speedup, demonstrating that \ta can efficiently process a $\approx 400$~GiB graph with only a $50$~GiB hot-store budget~(plus framework overheads).

\subsection{Impact of GPU Store Size}
\begin{figure}[t]
    \centering
    \subfloat[Inf. time and \# reloads.\label{subfig:gpu-1}]{
        \includegraphics[width=0.34\columnwidth]{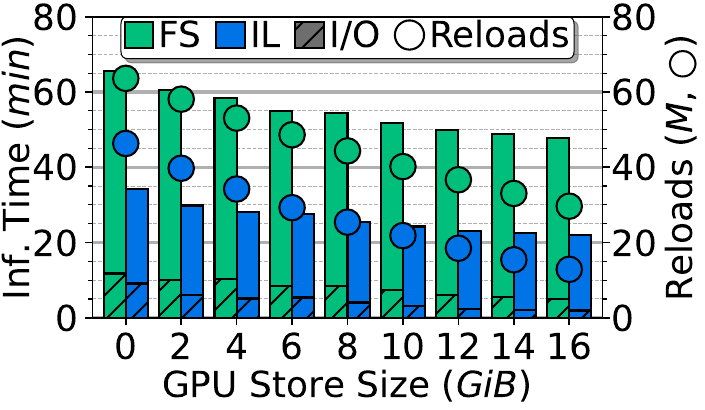}
    }\qquad\qquad
    \subfloat[\% of messages and vertices.\label{subfig:gpu-2}]{
        \includegraphics[width=0.34\columnwidth]{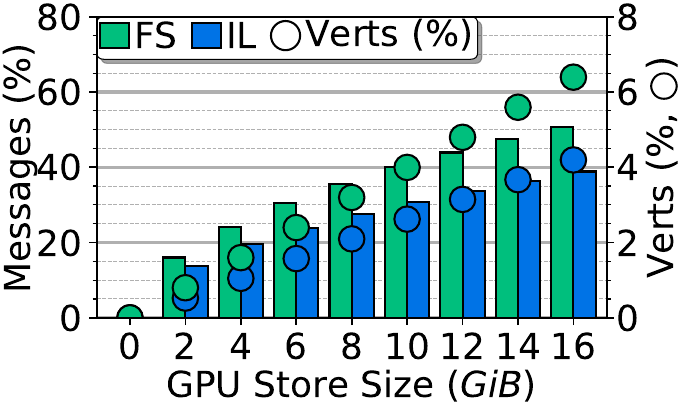}
    }
    \caption{Impact of GPU-store capacity on \ta. (a) E2E inference time~(\textit{solid bars}, left Y axis), I/O time~(\textit{hatched bars}, left Y axis) with corresponding cold-store reloads (\textit{markers}, right Y axis); (b) \% of messages aggregated~(\textit{bars}, left Y axis) and \% of vertices~(\textit{markers}, right Y axis) on the GPU-store.}
    \label{fig:gpustore}
\end{figure}

Fig.~\ref{fig:gpustore} evaluates the impact of GPU-store capacity on 2-layer \textit{full-graph} GCN inference for FS and IL using $80$~GiB and $40$~GiB hot stores, respectively.

\textit{Pinning hub vertices reduces host-memory pressure.}
Without a GPU store, all intermediate aggregation states are managed by the hot store, causing high-degree vertices to occupy hot-store slots for extended durations and repeatedly evict lower-degree vertices. At $0$~GiB, this results in $\approx64\M$ and $46\M$ cold-store reloads~(Figs.~\ref{subfig:gpu-1}, \textit{markers}, right Y axis) on FS and IL, respectively, with end-to-end inference times of $\approx66$ min and $35$~min and I/O times of $\approx12$ and $9$~min~(Figs.~\ref{subfig:gpu-1}, \textit{bars}, left Y axis). Increasing the GPU store permanently pins the highest-degree vertices, allowing them to directly absorb a disproportionate fraction of destination messages. 
At $16$~GiB, the GPU store caches just $6.4\%$ and $4.2\%$ of vertices on FS and IL~(Figs.~\ref{subfig:gpu-2}, \textit{markers}, right Y axis), yet serves $\approx51\%$ and $39\%$ of destination messages~(Figs.~\ref{subfig:gpu-2}, \textit{bars}, left Y axis), reducing reloads to $29.6\M$ and $12.9\M$ and inference time to $47$ and $22$~min.

\textit{GPU-store gains track SSD I/O and saturate quickly.} Fig.~\ref{subfig:gpu-1} shows that inference time closely tracks SSD I/O time as GPU-store capacity increases. At $16$~GiB, inference time reduces by $\approx27\%$~(FS) and $36\%$~(IL), while SSD I/O time decreases by $\approx57\%$ and $79\%$, respectively. The marginal benefit diminishes as capacity grows, since the highest in-degree vertices are pinned first. On IL, the first $2$~GiB reduces inference time by $\approx4.5$~min~($36\%$ of the total $12.3$~min gain), while the final $2$~GiB saves only $\approx0.5$~min. FS exhibits a similar pattern~($\approx5$ vs. $0.9$~min). Consequently, the first $8$~GiB captures $64$--$72\%$ of the total improvement on both datasets.

\begin{figure}[t]
  \centering
    \includegraphics[width=0.68\columnwidth]{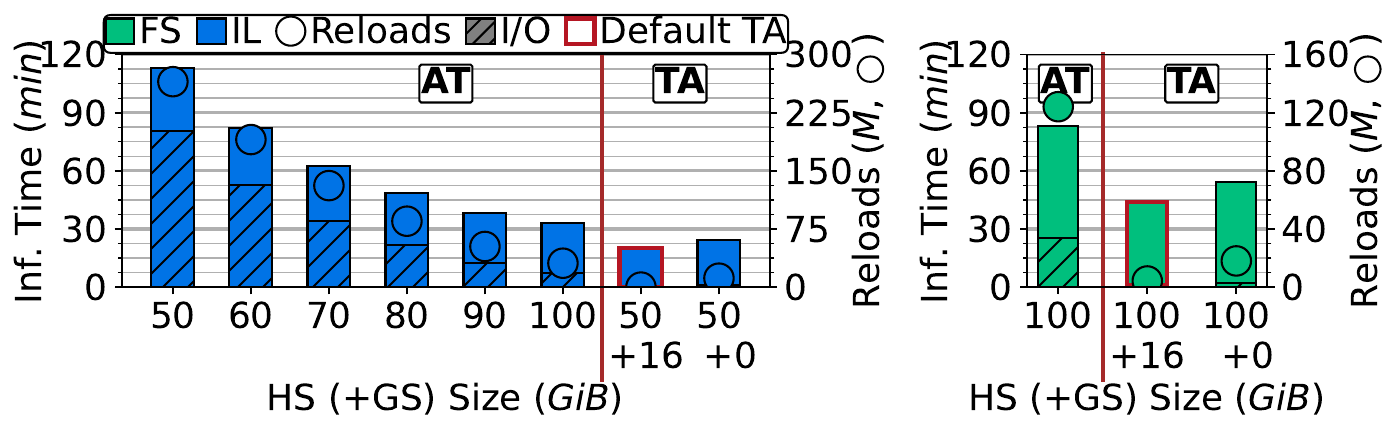}%
    \caption{\ta~(TA) vs. \at~(AT). E2E time~(\textit{bars}, left Y axis), I/O time~(\textit{hatched}), and \# reloads~(\textit{circles}, right Y axis) for varying HS~(+GS) budgets on IL~(\textit{blue}, left) and FS~(\textit{green}, right).}%
    \label{fig:atlas-vs-taurus}
\end{figure}

\subsection{\ta Reordering Cost}
\ta's reordering is a one-time preprocessing step and can be reused across all subsequent inference runs.
The RCMK bootstrap takes $\approx150$--$720$s~(largest on IF), after which the optimization loop itself is lightweight, with $5$ iterations taking $56$--$177$s~($5$--$15\%$) of total time. The dominant cost is feature relabeling~($125$--$1053$s, $23$--$80\%$), an I/O-bound operation, while topology relabeling adds only $35$--$111$s. Overall, preprocessing completes in $9$--$30$ min across all datasets.

\subsection{Comparison with \at}
Fig.~\ref{fig:atlas-vs-taurus} compares TA and AT for exact 2-layer GCN inference on IL and FS across hot-store~(HS) and GPU-store~(GS) budgets.
On IL, the default TA setup~($50{+}16$~GiB) completes in $\approx20$~min, $1.65\times$ faster than AT's \textit{best} at $100$~GiB HS~($33$~min) while using two-thirds of total memory, and $5.7\times$ faster than AT at an equal $50$~GiB HS. Even without a GPU store~($50{+}0$), matching AT's HS-only setup, \ta completes in $24.4$~min, $4.6\times$ faster than AT at the same memory budget, isolating the benefit of \ta's topology-aware reordering. TA's I/O is negligible~(\textit{hatched}, $1.3$~min vs. $7$--$81$~min for AT) as reloads fall from $265\M$ to $11.5\M$~(\textit{circles}). On FS at an equal $100$~GiB HS, TA is $1.9\times$ faster~($43.9$ vs. $83.1$~min), and $1.5\times$ faster even with no GS. We omit additional AT memory configurations on FS due to prohibitively long runtimes. 
Overall, TA outperforms AT with lower memory and substantially lower I/O.

\subsection{Resource Utilization of \ta}

\begin{figure}[t]
    \centering
    \subfloat[FS/GCN2 CPU and Memory\label{subfig:fs-cpu-mem}]{
        \includegraphics[width=0.44\linewidth]{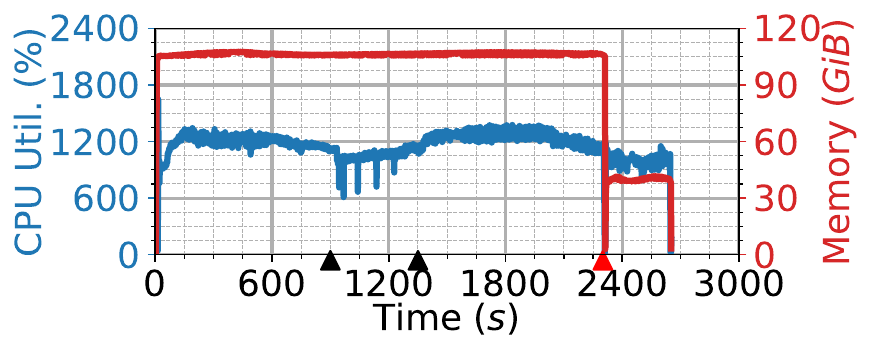}
    }
    \subfloat[FS/GCN2 GPU and VRAM\label{subfig:fs-gpu-gmem}]{
        \includegraphics[width=0.44\linewidth]{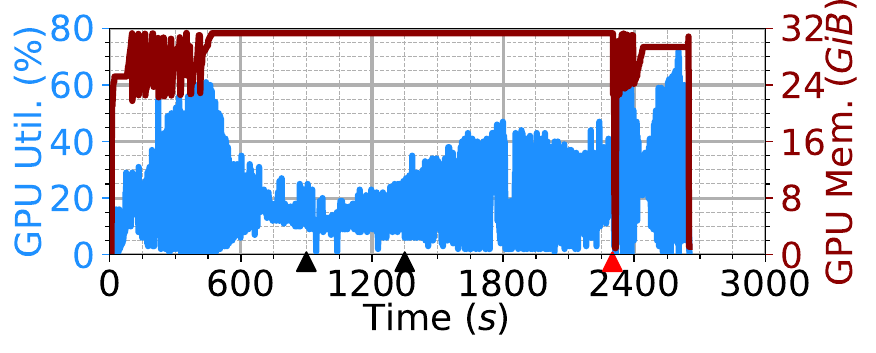}
    }\\
    \subfloat[FS/GCN2 Read/Write Bandwidth\label{subfig:fs-rw}]{
        \includegraphics[width=0.44\linewidth]{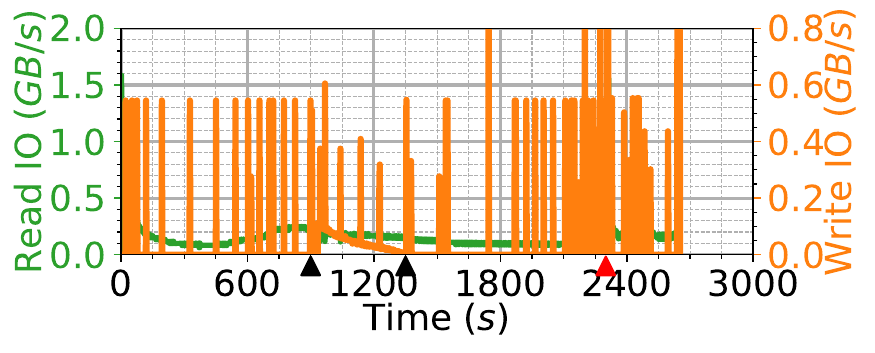}
    }
    \caption{Resource util. for a 2-layer GCN on the FS dataset. (a) CPU~(\textit{blue}, left Y axis) and memory~(\textit{red}, right Y axis) over time; (b) GPU util~(\textit{sky blue}, left Y axis) and memory~(\textit{brick}, right Y axis) usage over time;
    (c) SSD read~(\textit{green}, left Y axis) and write~(\textit{orange}, right Y axis) bandwidth over time.}
    \label{fig:util}
\end{figure}
Fig.~\ref{fig:util} shows CPU utilization, RAM usage, GPU utilization and memory, and SSD read/write throughput, sampled every second during exact 2-layer GCN inference on FS with a 100~GiB hot store and a 16~GiB GPU store.

\ta sustains average $1150\%$ CPU utilization~(Fig.~\ref{subfig:fs-cpu-mem}, \textit{blue lines}, left Y axis) on our 12-core~(2-way hyperthreaded) machine~(\S~\ref{subsec:eval-setup}) while average SSD reads remain low at $160$~MiB/s (Fig.~\ref{subfig:fs-rw}, \textit{green lines}, left Y axis), hence the streaming readers and writers do not stall compute, with an initial read burst~($\approx1.8$~GiB/s) at layer starts reflecting filling of the read queue. CPU utilization drops ($1200\rightarrow1050\%$) while SSD writes increase between $900$--$1350$ s~(\textit{black arrows}) due to vertex evictions.
Resident memory~(Fig.~\ref{subfig:fs-cpu-mem}, \textit{red lines}, right Y axis) tracks the configured budget $\approx105$~GiB in Layer~1~($100$~GiB hot store plus $\approx5$~GiB framework overhead) and $\approx38$~GiB in Layer~2 once $128$-dim embeddings replace $1024$-dim input features with the step at $\approx2700$~s marking the layer transition~(\textit{red arrow}).
GPU utilization~(Fig.~\ref{subfig:fs-gpu-gmem}, \textit{sky blue lines}, left Y axis) is intermittent~(peak $72\%$) because CPU-side aggregation and memory management dominate. GPU memory~(Fig.~\ref{subfig:fs-gpu-gmem}, \textit{brick lines}, right Y axis) stays around $\approx31$~GiB reflecting the $16$~GiB GPU store, $8$~GiB write buffers, plus transient buffers.
The periodic write spikes~(peak $4.3$~GiB/s) mark graduated vertices being flushed to the SSD.

\section{Related Works}\label{sec:related}
GNNs support large-scale recommendation and traffic applications~\cite{pinsage,aligraph,derrow2021eta}. As graphs evolve, systems periodically retrain models and refresh vertex representations to address drift in topology, features and labels~\cite{rossi2020temporal,xia2023redundancy}. \ta targets this refresh phase, accelerating billion-scale inference on a single OOC machine.

\subsection{Large-scale GNN Training and Inference}
While many systems target scalable GNN \textit{training}~\cite{p3,aligraph,distdgl}, comparatively little attention has been paid to large-scale \textit{inference}. Systems such as AliGraph~\cite{aligraph} and DistDGL~\cite{distdgl} are designed for distributed mini-batch training, partitioning graphs across machines and optimizing sampling and gradient synchronization. Likewise, popular GNN libraries such as PyG~\cite{fey2019fast} and DGL~\cite{wang2019deep} provide extensive support for mini-batch training and neighborhood sampling, but are not optimized for single-machine full-graph inference. Consequently, billion-scale inference typically relies on distributed deployments~(e.g., DistDGL-style clusters), incurring significant infrastructure overheads. 
Exact full-neighborhood inference creates larger working sets than sampled training, while sampled inference still suffers repeated feature movement under gather-based execution.
Recent systems target GNN inference more directly. InferTurbo~\cite{inferturbo} uses GAS-style execution for scalable inference, but relies on large MapReduce clusters. DGI~\cite{dgi} converts training code to layer-wise inference with dynamic batching and graph reordering; however, its mmap-based OOC design remains gather-driven and incurs high read amplification beyond RAM. We therefore use DGI as the strongest layer-wise baseline.

\subsection{Out-of-core GNN Training Systems}
Several systems use SSD-backed single machines for large-scale OOC GNN training~\cite{capsule,outre,marius,diskgnn,ginex,hyperion,caliex}, primarily optimizing data movement, storage layout, and caching.
DiskGNN~\cite{diskgnn} reduces read amplification by precomputing computation graphs, packing features contiguously on disk, and pipelining execution over a multi-level feature store, while MariusGNN~\cite{marius} avoids precomputation by partitioning the graph and restricting aggregation to a vertex's memory-resident neighbors within a partition.
These strategies do not directly fit exact full-neighborhood inference: all vertices require outputs, precomputing every computation graph is prohibitive, and partition-restricted aggregation can drop cross-partition neighbors~\cite{diskgnn}.
Capsule~\cite{capsule} uses partitioning, pruning, and optimized loading to fit training subgraphs in GPU memory, while Ginex~\cite{ginex} separates sampling from feature gathering to enable Belady-optimal feature caching. These optimizations target sampled training; inference over all vertices, exact or sampled, still stresses feature movement and exposes random gather overheads.

In contrast, \ta makes sequential source broadcasts the primary abstraction for inference, then manages the resulting partial states through topology-aware ordering and tiered GPU--RAM--SSD storage.

\section{Conclusion and Discussion}\label{sec:discussion}
We introduced \ta, a single-machine out-of-core system for billion-scale GNN inference, supporting both exact full-neighborhood and fanout-sampled execution.
By replacing destination-centric gathers with pipelined source broadcasts, \ta converts repeated random reads into sequential scans.
Coupled with a tiered GPU-CPU-SSD hierarchy, topology-aware vertex reordering, and a pending-message eviction policy, it minimizes hot-store residency and I/O thrashing. Despite these gains, \ta currently assumes static graph snapshots. We plan to address this by extending the broadcast execution model to support continuous, incremental updates on evolving topologies, and we also plan to explore multi-GPU scaling to accelerate compute-bound phases and support even larger embedding dimensions.

\section*{Acknowledgments}
The authors thank Roopkatha Banerjee and other members of the DREAM:Lab, Indian Institute of Science, for their assistance and insightful feedback. 
The authors used AI-assisted tools only for language refinement and clarity. All ideas, methods, experiments, results and conclusions are the authors' own, and the final manuscript was reviewed and approved by the authors.

\bibliographystyle{IEEEtran}
\bibliography{refs}
\end{document}